\documentclass[fleqn,usenatbib]{mnras}

\usepackage{mathptmx}

\usepackage[T1]{fontenc}

\DeclareRobustCommand{\VAN}[3]{#2}
\let\VANthebibliography\thebibliography
\def\thebibliography{\DeclareRobustCommand{\VAN}[3]{##3}\VANthebibliography}

\usepackage{graphicx}	
\usepackage{amsmath}	
\usepackage{amssymb}	

\usepackage{bm}
\usepackage{color}
\usepackage{ulem}
\usepackage{xspace}
\newcommand{\Se}[1]{Section~\ref{sec:#1}}
\newcommand{\se}[1]{Sect.~\ref{sec:#1}}
\newcommand{\Fg}[1]{Fig.~\ref{fig:#1}}
\newcommand{\fg}[1]{Fig.~\ref{fig:#1}}
\newcommand{\fgs}[2]{Figs.~\ref{fig:#1} and \ref{fig:#2}}
\newcommand{\tb}[1]{Table~\ref{tab:#1}}
\newcommand{\Tb}[1]{Table~\ref{tab:#1}}
\newcommand{\eq}[1]{Eq.~(\ref{eq:#1})}
\newcommand{\Eq}[1]{Equation~(\ref{eq:#1})}
\newcommand{\Eqs}[2]{Equation~(\ref{eq:#1}) and (\ref{eq:#2})}

\title[Large Pebble Accretion]{Accretion of aerodynamically large pebbles}

\author[Huang and Ormel]{
Helong Huang$^{1}$, 
Chris W. Ormel$^{1}$ \thanks{E-mail: chrisormel@tsinghua.edu.cn}
\\
$^{1}$Department of Astronomy, Tsinghua University, Haidian DS 100084, Beijing, China
}

\date{Accepted XXX. Received YYY; in original form ZZZ}

\pubyear{2023}

\begin{document}
\label{firstpage}
\pagerange{\pageref{firstpage}--\pageref{lastpage}}
\maketitle

\begin{abstract}
    Due to their aerodynamical coupling with gas, pebbles in protoplanetary discs can drift over large distances to support planet growth in the inner disc. In the past decade, this pebble accretion has been studied extensively for aerodynamically small pebbles (Stokes number $\mathrm{St}<1$). 
    However, accretion can also operate in the $\mathrm{St}>1$ mode, e.g., when planetesimals collisionally fragment to smaller bodies or when the primordial gas disc disperses. 
    This work aims to extend the study of pebble accretion to these aerodynamically loosely coupled particles. 
    We integrate the pebble's equation of motion, accounting for gas drag, stellar and planetary gravity, in the midplane of a laminar disc.
    The accretion probability ($\epsilon$) is calculated as function of Stokes number, disc pressure gradient index, planet mass and eccentricity.
    We find that for Stokes number above unity $\epsilon(\mathrm{St})$ first rises, due to lower drift and aided by a large atmospheric capture radius, until it reaches a plateau where the efficiency approaches $100$ per cent. At high St the plateau region terminates as particles become trapped in resonance.
    These results are well described by a semi-analytical ``kick-and-drift'' model and we also provide fully analytical prescriptions for $\epsilon$.
    We apply our model to the accretion of ${\sim}30\,\mu\mathrm{m}$ dust particles in a dispersing protoplanetary and secondary (CO-rich) debris disc. It shows that such physically small particles are mainly accreted as aerodynamically large Stokes number pebbles during the debris disc phase. Earth-mass planets may obtain $\sim$25 per cent of their heavy elements through this late accretion phase.
\end{abstract}




\section{Introduction}
Protoplanetary discs around young stellar objects are found to emit at millimetre and submillimetre wavelengths \citep{LooneyEtal2000, AndrewsWilliams2005, ALMA2015, AnsdellEtal2016}.  Most prominently, the Atacama Large Millimeter/submillimeter Array (ALMA) has revealed the prevalence of disc substructure, like rings, gaps, spiral arms and clumps in the last decade \citep{ALMA2015, AndrewsEtal2018, DongEtal2018, PerezEtal2018, Andrews2020}. These observations shed light on the fundamental physical processes governing disc and planet formation. 
The leading interpretation is that the particles responsible for the continuum emission are `pebbles' -- particles that have grown in the disc out of (sub)micron-sized grains by coagulation \citep{Smoluchowski1916, JohansenEtal2014,DrazkowskaEtal2022, MiotelloEtal2022}. The physical size of these particles is debated and ranges from ${\sim}100\,\mu\mathrm{m}$ as inferred from polarization studies \citep{KataokaEtal2015, BacciottiEtal2018} to ${\sim}\mathrm{cm}$ as inferred from multi-wavelength analysis \citep{PerezEtal2012,TazzariEtal2016,KimEtal2019,MaciasEtal2021}. The total mass budget of pebbles in young, Class I disks varies and ranges between $M_\oplus$ and $10^3 M_\oplus$ \citep{TychoniecEtal2018,TychoniecEtal2020}, while the mass budget for more evolved, Class II disk ranges from $0.1 M_\oplus$ to $100M_\oplus$ \citep{AnsdellEtal2016,AnsdellEtal2017}. At early times, it is likely that they dominated the solid mass budget.

Pebbles\footnote{Throughout this work, following the precedence which has emerged in the astrophysical liturature, we refer to `pebbles' in an aerodynamical context. A physically small grain can therefore be an (aerodynamically) large ($\mathrm{St}>1$) pebble at the same time.} are (by definition) aerodynamically active, and tend to move to regions of high pressure \citep{WeidenSchilling1977}.  For smooth disks, where the pressure gradient is radially negative, this results in their inward radial drift \citep{WeidenSchilling1977}.  However, a local reversal of the pressure distribution, e.g, due to gap opening by planets \citep{LinPapaloizou1986,Rafikov2002,ZhuEtal2014} or differential accretion flux at the edge of an MRI dead zone \citep{VarniereTagger2006, FlockEtal2015}, will likewise show an imprint on the distribution of pebbles in discs.
Pebbles are also important in driving planet growth. First, they are thought to make up a large fraction, if not dominate, the total solid mass budget in protoplanetary discs. But the main reason why pebbles are ideal for planet formation is their mobility -- the pebble reservoir in the outer disc may drift inwards to aid planet growth in the inner disc regions. 
When a pebble drifts across the planet orbit, the gravity from the planet will attract the pebbles towards it. Without gas drag, most of these encounters will merely result in their gravitational scattering rather than accretion \citep{GoldreichEtal2004a}. However, for pebbles gas drag dissipates their energy effectively \textit{during} the encounter, which results in its settling and eventual capture by the planet.  This process is called `pebble accretion', first described by \citet{OrmelKlahr2010}.  

The aerodynamical property of pebbles is quantified by their Stokes number (St, see \se{eom}). Small pebbles ($\mathrm{St}<1$), couple to the gas on time-scales much shorter than the orbital time-scales. The capture properties of $\mathrm{St}<1$ particles depend on the Stokes number, the properties of the disk, and the planet mass, but is independent of the planet physical radius \citep{Ormel2017}. In \fg{sketch-settling-ballistic} this is illustrated by the sharp division between ``settling encounters'' (pebble accretion) and ballistic encounters: for $\mathrm{St}<1$ the lines overlap.  \citet{LiuOrmel2018} and \citet{OrmelLiu2018} derived closed-form analytical formula for pebble accretion, accounting for planet eccentricity, inclination and disc turbulence.  \citet{LambrechtsJohansen2012} and \citet{BitschEtal2015} found that the growth time-scale of giant planet cores by pebble accretion is shortened significantly compared to planetesimal accretion \citep{JohansenBitsch2019}, especially in the outer disc.
Another environment favourable to pebble accretion are locations where the radial and azimuthal drift of pebbles is greatly reduced, as a result of particle pile-ups or pressure traps, such as the continuum rings observed by ALMA \citep{JiangOrmel2022, LauEtal2022}.
Pebble accretion has been applied towards the formation of ice giants \citep{LambrechtsJohansen2014}, gas giants \citep{BitschEtal2015, WimarssonEtal2020, GuileraEtal2020} and terrestrial planets \citep{OrmelEtal2017, LambrechtsEtal2019, SchoonenbergEtal2019} and is also a key component in planet population synthesis models \citep{Chambers2016, LiuEtal2019, VenturiniEtal2020}. Pebble accretion is successful in matching the properties of planets in the solar system, in reproducing the masses, orbits and composition of terrestrial planets \citep{JohansenEtal2021} and the metallicity of giant planets \citep{LambrechtsEtal2014}.

Planetesimals are defined as bodies of size ${\gtrsim}1\,\mathrm{km}$ sizes, so the gas drag they feel is negligible compared to the gravitational force. When the relative velocity between planetesimals is low, planetesimals accretion is in the runaway growth regime, because the gravitational focusing becomes stronger for larger planetesimals \citep{WetherillStewart1989,KokuboIda1996}. However, when the largest planet embryo becomes massive enough to gravitationally stir up the nearby planetesimals, planetesimal accretion slows down, entering the oligarchic growth regime, where gravitational focusing is suppressed by increasingly higher approach velocities \citep{IdaMakino1993,KobukoIda1998}. As a result, planetesimal accretion at large distances suffers from a long formation time-scale, especially outside of the snow line \citep{IdaLin2004}. 
It is found that a giant planet core at a distant orbit ($\gtrsim 10\ \mathrm{au}$) can hardly reach $10\ M_\oplus$ by accreting planetesimals \citep{DodsonRobinsonEtal2009,KobayashiEtal2010}, which is the canonical threshold to trigger runaway gas accretion. 
In addition, planetesimals may become trapped in a orbital mean motion resonances (MMR) with the planet, preventing their accretion \citep{WeidenschillingDavis1985,TanakaIda1999,HahnMalhotra1999,MustillWyatt2011,ShibataEtal2020}.
In conclusion, planetesimal accretion alone cannot explain the formation of planets at large distances.

\begin{figure}
        \includegraphics[width=\columnwidth]{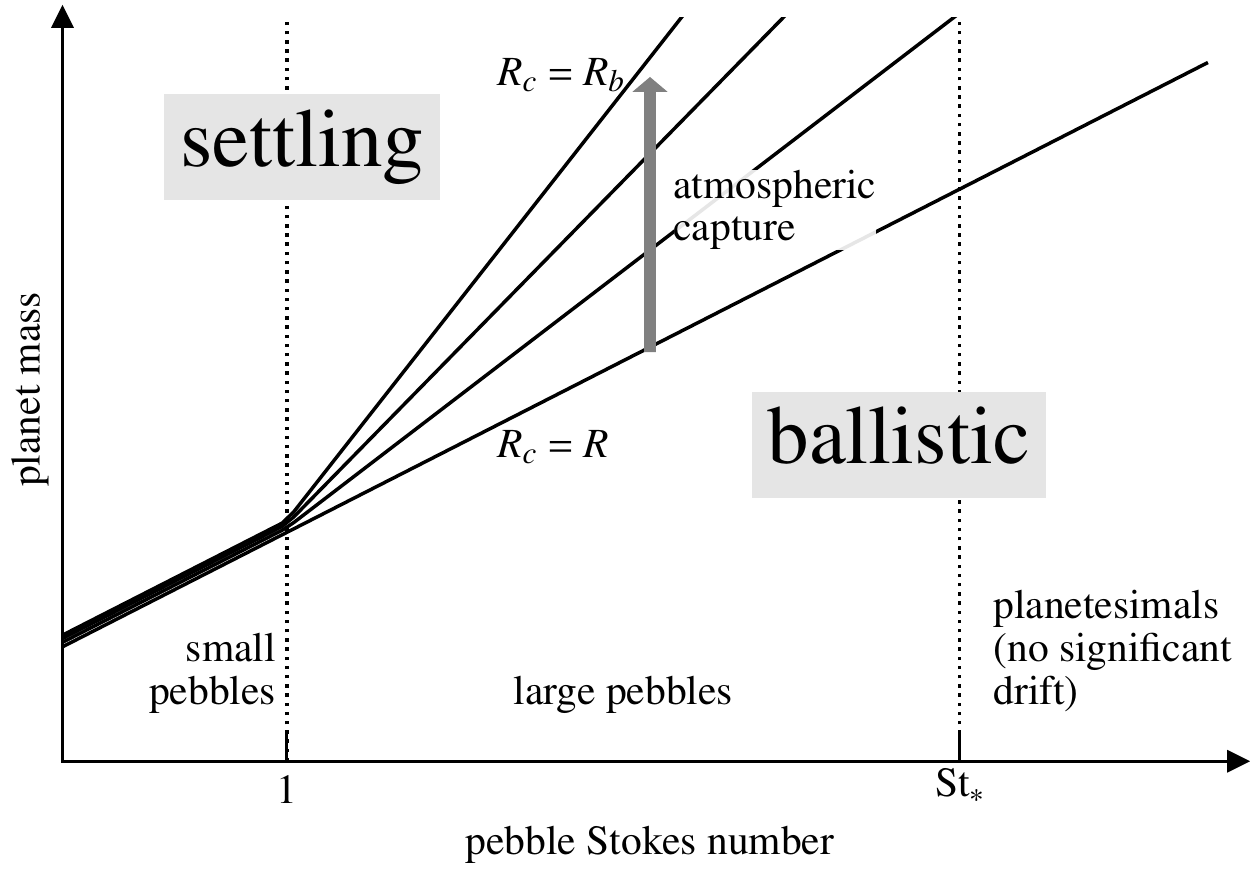}
        \caption{\label{fig:sketch-settling-ballistic} 
        Nature of pebble accretion for small and large Stokes numbers. The lines schematically denote the division where settling interactions (accretion independent of the planet physical size) and ballistic interactions (accretion depends on hitting a surface) dominate. We define \textit{pebbles} as any particle for which drift is signifcant on evolutionary time-scales, $\mathrm{St}<\mathrm{St}_\ast$. Large pebbles (not necessary large particles) are those with Stokes number $1<\mathrm{St}<\mathrm{St}_\ast$. For small pebbles ($\mathrm{St}<1$) pebble accretion occurs through the settling mechanism, with capture occuring under nebula conditions, independent of the planet physical size. For pebbles of $\mathrm{St}>1$ nebular drag tends to become increasingly less effective and it is more likely that these large pebbles will be accreted ballistically after entering the protoplanet atmosphere, whose radius $R_c$ could exceed up to the Bondi radius $R_b$. }
\end{figure}

On the other hand, accretion of planetesimals by planets is boosted by the bloated nature of the pre-planetary atmosphere that form after the protoplanet has become sufficiently massive. These atmospheres act to enhance the effective capture radius of these planets \citep{InabaIkoma2003}. 
This `atmospheric enhancement' is similar in nature to pebble accretion as both mechanisms rely on energy dissipation by gas drag. But whereas for $\mathrm{St}<1$ pebbles capture already happens under nebula conditions, the atmospherically-induced capture requires an increase in gas density associated with the planet's proto-atmosphere. We therefore classify these interactions as ballistic (see \fg{sketch-settling-ballistic}). Recently, \citet{OkamuraKobayashi2021} studied pebble accretion with a wide size range (Stokes number), also accounting for the gas flow perturbed by the planet.  They found that large pebbles get captured efficiently due to their supersonic velocity and that proto-atmospheres can significantly enhance the accretion rate.  However, they studied the trajectory of pebbles in a local frame (similar to \citealt{OrmelKlahr2010}), which cannot by definition account for global effects like repeated encounters by planets or trapping of planets in mean motion resonances.

This work focuses on the accretion of \textit{aerodynamically large} pebbles, which we define as having dimensionless stopping time (or Stokes number) $\mathrm{St}\gtrsim 1$ [see Eq.(\ref{eq:St_Ep}) -- (\ref{eq:St})] but still small enough for them to drift over significant distances (e.g., the planet's feeding zone) over evolutionary time-scales. 
Following \citet{LiuOrmel2018}, we carry out numerical planet-pebble integrations in the global frame, which allows them to be captured in mean motion resonances and prevents their double counting. The outcome of each individual encounter is then recorded and this allows us to obtain the pebble accretion efficiency $\epsilon$ -- the fraction of the total pebble mass flux that is accreted by the planet.

The structure of this paper is as follows. In \Se{model} we present the physical model to calculate the pebbles' orbits and the accretion efficiency. \Se{result} discusses the outcome of numerical simulation and \Se{parameter} demonstrates the parameter study in which we change disc pressure profile, planet mass and eccentricity. We present a semi-analytical model and analytical fit formulae for $\epsilon$ in \Se{semiana} and \ref{sec:analytical}.  Then in \Se{application} we apply our findings to a planet in a disc undergoing photoevaporation, showing the potential to accrete aerodynamically large pebbles in debris discs. We compare this work with previous studies on pebble accretion and state the caveat of our work in \Se{discussion}.  Finally we summarize our results in \Se{conclusion}.
\begin{table*}
    \caption{Simulation parameters}
    \label{tab:parameters}
    \centering
    \small
    \begin{tabular}{l l l l l}
        \hline
        drag regime & planet-to-star mass ratio $q$ & pressure gradient $\eta$ & planet eccentricity $e_p$ & reference\\
        \hline
        Epstein & $3\times 10^{-6}$ & $3\times 10^{-3}$ & $0$ & \fg{default}\\
        Epstein & ($3\times 10^{-7}$, $3\times 10^{-6.9}$, \dots, $3\times 10^{-5}$) & $3\times 10^{-3}$ & $0$ & \fg{q}\\
        Epstein & $3\times 10^{-6}$ & ($10^{-3.5}$, $10^{-3.4}$, \dots, $10^{-2}$) & $0$ & \fg{eta}\\
        Epstein & $3\times 10^{-6}$ & $3\times 10^{-3}$ & ($0$, $10^{-3}$, $10^{-2.9}$, \dots, $10^{-1}$) & \fg{e}\\
        Stokes & $3\times 10^{-6}$ & $3\times 10^{-3}$ & $0$ & \fg{app-default}\\
        Stokes & ($3\times 10^{-7}$, $3\times 10^{-6.9}$, \dots, $3\times 10^{-5}$) & $3\times 10^{-3}$ & $0$ & \fg{app-q}\\
        Stokes & $3\times 10^{-6}$ & ($10^{-3.5}$, $10^{-3.4}$, \dots, $10^{-2}$) & $0$ & \fg{app-eta}\\
        Stokes & $3\times 10^{-6}$ & $3\times 10^{-3}$ & ($0$, $10^{-3}$, $10^{-2.9}$, \dots, $10^{-1}$) & \fg{app-e}\\
        \hline
    \end{tabular}
\end{table*}

\section{Model} \label{sec:model}
\subsection{Equation of motion} \label{sec:eom}

We consider a planet on a Keplerian orbit embeded in a laminar photoplanetary disc. For simplicity, we assume that the planet is small enough not to affect the gas properties so that the gas motion is in the azimuthal direction, $v_\mathrm{gas,\phi} = (1-\eta)v_\mathrm{k}$. Here, $v_\mathrm{k}$ is the Keplerian velocity at radius $r$ and 
$\eta$ is the pressure gradient parameter, defined as $\eta \equiv - (c_s^2/2v_\mathrm{k}^2) (\mathrm{d} \log P/\mathrm{d} \log r)$, where $c_s$ is the local isothermal sound speed. $\eta$ describes the pressure support of the gas disc, resulting in its sub-Keplerian motion.  Pebbles in the disc feel gas drag, losing angular momentum and drifting inward, allowing them to be captured by the planet.

The gas drag a solid body feels in solar nebular is summarized by \citet{WeidenSchilling1977}. When the particle radius $s<9\lambda/4$, where $\lambda$ is the mean free path of the gas, drag obeys the Epstein regime,
\begin{equation}
    \label{eq:St_Ep}
    F_{\mathrm{drag, Ep}} = \frac{4\pi}{3}\sqrt{\frac{8}{\pi}}c_s s^2 \rho_g v.
\end{equation}
Here, $\rho_g$ is gas density and $v$ is the relative velocity between gas and particle. When $s>9\lambda/4$, the drag obeys Stokes regime. In the main text we only adopt Epstein drag law, which applies for small particles or for dilute gas medium, e.g., in the outer protoplanetary discs or debris discs. The simulation involving Stokes drag law is shown in Appendix. The gas drag force can equivalently be described in terms of a stopping time 
\begin{equation}
    t_s = \frac{mv}{F_{\mathrm{drag}}},
    \label{eq:drag}
\end{equation}
where $m$ is the particle mass. We further normalise the stopping time with $\Omega_\mathrm{k}$, the circular Keplerian angular velocity at the pebble's location, to define the dimensionless quantity Stokes number,
\begin{equation}
    \mathrm{St}=t_s\Omega_\mathrm{k}.
    \label{eq:St}
\end{equation}
Thus we conclude that in the Epstein regime, $\mathrm{St}$ is independent of the relative velocity between pebble and gas. 
This work treats two gas drag regime seperately, that is, a single simulation setup describes either the Epstein regime or the Stokes regime, see \Tb{parameters}. 

The equation of motion for a pebble in a coordinate frame centred on the star is
\begin{equation}
    \frac{\mathrm{d}^2 \bm{r}}{\mathrm{d} t^2} = -\frac{GM_\ast\bm{r}}{r^3} - \frac{GM_p(\bm{r}-\bm{r}_p)}{|\bm{r}-\bm{r}_p|^3} - \frac{1}{t_s}(\bm{v}-\bm{v}_\mathrm{gas}) - \frac{GM_p\bm{r}_p}{r_p^3}.
\end{equation}
The first and second terms on the right hand side describe the gravity from star and planet, respectively, where $\bm{r}$ and $\bm{r}_p$ are the position vector of the pebble and the planet relative to the star, with norms $r$ and $r_p$, and $M_\ast$ and $M_p$ are the stellar mass and planet mass. The third term is the gas drag force. The last term accounts for the acceleration of the coordinate system due to the planet's gravity (indirect force).

To generalise the results, we use dimensionless units where lengths are normalised by the planet semi-major axis $a_p$, mass by $M_\ast$ and time by the planet orbital period $\Omega_0^{-1}$. In these units, the gravitational constant $G=1$ and the equation of motion becomes
\begin{equation}
    \frac{\mathrm{d}^2 \bm{r'}}{\mathrm{d} {t'}^2} = -\frac{\bm{r'}}{{r'}^3} - q\frac{(\bm{r}'-\bm{r}'_p)}{|\bm{r}'-\bm{r}'_p|^3} - \frac{1}{\mathrm{St}}\frac{\bm{v'}-\bm{v}'_\mathrm{gas}}{{r'}^\frac{3}{2}} - \frac{q\bm{r}'_p}{{r'_p}^3}.
    \label{eq:em_vec}
\end{equation}
Here $q=M_p/M_\ast$ is the planet-to-star mass ratio.

We only consider the accretion of the pebble in a 2D plane, so the planet does not have an inclination. The free parameters in our model are $q$, $\eta$, planet eccentricity $e_p$, $\mathrm{St}$ and the gas drag law. The advantage of using the above normalisation is that it reduces the number of free parameters. For example, any ($M_p$, $M_\ast$) pair amounting to the same $q$ will give the same pebble trajectories in dimensionless units. 
We fixed parameters in each simulations as listed in table \tb{parameters}.

\subsection{Numerical method}

Pebbles' 2D orbits are integrated in the rotational frame centred on the star using RKF45 method \citep{Fehlberg1969}, and their contribution to the planet's growth is studied. 
The numerical error, defined as the maximum error among all components of the dimensionless position and velocity vector within one time step $\max\{err(\mathrm{d}\bm{r'}), err(\mathrm{d}\bm{v'})\}$, is controlled to be smaller than $10^{-8}$ outside the Hill radius, $10^{-9}$ inside the Hill radius and $10^{-10}$ inside 10 per cent of the Hill radius. 
Because inside Hill radius $r_\mathrm{h} = a_p(q/3)^{1/3} \equiv a_pr'_\mathrm{h}$, the planet's gravity dominates and we need more precise description of the pebble's position and velocity.

For each parameter tuple $(\eta, q, e_p, \mathrm{St})$, we start N (N=2880 for default parameters and 1440 for parameter study) pebbles following a uniform distribution on the circle around the star, with radius $1.75a_p$, to ensure that every first order mean motion resonance is covered. The initial velocity of the pebbles is assumed to be the unperturbed drift velocity \citep{WeidenSchilling1977},
\begin{eqnarray}
    \begin{aligned}
    v'_\mathrm{r} &= -\frac{2\eta \mathrm{St}}{\mathrm{St}^2+1}\\
    v'_\mathrm{t} &= 1-\frac{\eta}{\mathrm{St}^2+1}.
    \label{eq:vdrift}
    \end{aligned}
\end{eqnarray}

Each simulation results in one of the following four outcomes:
\begin{enumerate}
    \item Settling -- the partice is permanently captured in the gravitational well of the planet. It settles towards the planets.
    \item Ballistic hit -- the particle is not captured by the planet, but ends up interior to the planet's orbit. The minimum distance to the planet is recorded, from which the balistic hit fraction is calculated \textit{a posteriori}.\footnote{We recorded the each pebble's minimum distance to the planet. By comparing them to the planet radius, we get the probability for ballistic hit with planets of different sizes.}
    \item Missing -- the particle ends up inside the planet orbit. The minimum distance to the planet is larger than the planet physical radius.
    \item Resonance -- the particle is captured in resonance exterior to the orbit of the planet.
\end{enumerate}

The criterion for settling is that the pebble completes $\mathrm{St}+25$ revolutions about the planet within $0.1r_\mathrm{h}$, as within $0.1r_\mathrm{h}$ planet gravity dominate over the solar stellar gravity and centrifugal force. To assess the sensitivity of the result to the threshhold, the critical revolution number of $\mathrm{St}/2+12$ (low threshold) is also tested. For most Stokes numbers ${>}1$, at most $10$ per cent of the settling encounters are instead counted as ballistic hits compared with the low threshold. This is due to the unrealistic deflection of pebbles caused by numerical precision errors when the captured pebble spirals in on an orbit which periastron distance is too close to the planet. However, as such trajectories have already satisfied the condition for a ballistic hit, these numerical errors bear no influence on the (total) accretion probability or any of our conclusions.
In missing encounters or ballistic hits the particle crosses the inner integration boundary set to be $(1-e_p)a_p - 5r_\mathrm{h}$, where $e_p$ is the planet eccentricity. 

In order to study the relation between the ballistic hit probability and the planet physical radius, we record the minimum distance of the pebble to the planet. In the protoplanetary discs planets will acquire a dense primordial (hydrogen-helium) atmosphere once their surface escape velocity exceeds the thermal velocity of the disc gas. When pebbles enter these `proto-atmospheres', their gas drag is elevated \citep{InabaIkoma2003}. Formally the effective capture radius can be computed numerically if the density structure of the atmosphere is known \citep{OkamuraKobayashi2021}. For simplicity, in most cases we take the Bondi radius $r_b=GM_p/c_s^2$ as the typical planet physical radius.
The Bondi radius signifies the point where the escape velocity exceeds the isothermal sound speed, so that the gas density significantly increases and the pebble is likely to be captured. The approximation to equate the effective physical radius with $r_b$ is better for low-St particles because of stronger drag \citep{InabaIkoma2003}. If we define the disc aspect ratio to be $h = c_s/v_\mathrm{k}$, then in dimensionless units, $r'_b=q/h^2$. In our simulation, we take the disc aspect ratio $h=0.05$.

For the resonant outcome, we examined the averaged radial velocity $\overline{dr'/dt'}$, over $n_\mathrm{crit}$ encounters with the planet.  In the absence of a planet, this rate should be the unperturbed drift velocity (\eq{vdrift}).  We set the threshold to $10^{-4}v'_r$. When the averaged radial velocity is smaller than this threshold, we consider that the pebble is trapped in resonance. The critical number $n_{crit}$ is not kept constant but self-adjusts in the simulation to account for pebbles jumping from different resonant orbits. 
The initial value of $n_\mathrm{crit}$ is set to be 10. Once the averaged drift velocity is less than $0.05v'_r$, we double $n_\mathrm{crit}$ until the criterion is satisfied. We classify these pebbles to be in resonance.

However, to avoid some extreme cases when $\mathrm{St}$ is very large, we set the time-out of the integration to be $10^6$ in dimensionless unit, which corresponds to $1.6\times 10^5$ yr in Earth orbit and $1.9\times 10^6$ yr for Jupiter.  If time-out is reached, the pebble is also classified to be in resonance, because its accretion time-scale is longer than the time-out.

The settling efficiency $\epsilon_\mathrm{set}$ and ballistic hit efficiency $\epsilon_\mathrm{bal}$ are defined as the fraction of pebbles settling to the planet and ballistically hit with the planet. 
The accretion efficiency is defined as the combination of $\epsilon_\mathrm{set}$ and $\epsilon_\mathrm{bal}$,
\begin{equation}
    \epsilon = \frac{n_\mathrm{settling}+n_\mathrm{hit}}{n_\mathrm{settling}+n_\mathrm{hit}+n_\mathrm{miss}+n_\mathrm{resonance}}.
\end{equation}

\section{Results}
\subsection{Numerical result for default runs} \label{sec:result}

\begin{figure*}
        \includegraphics[width=\textwidth]{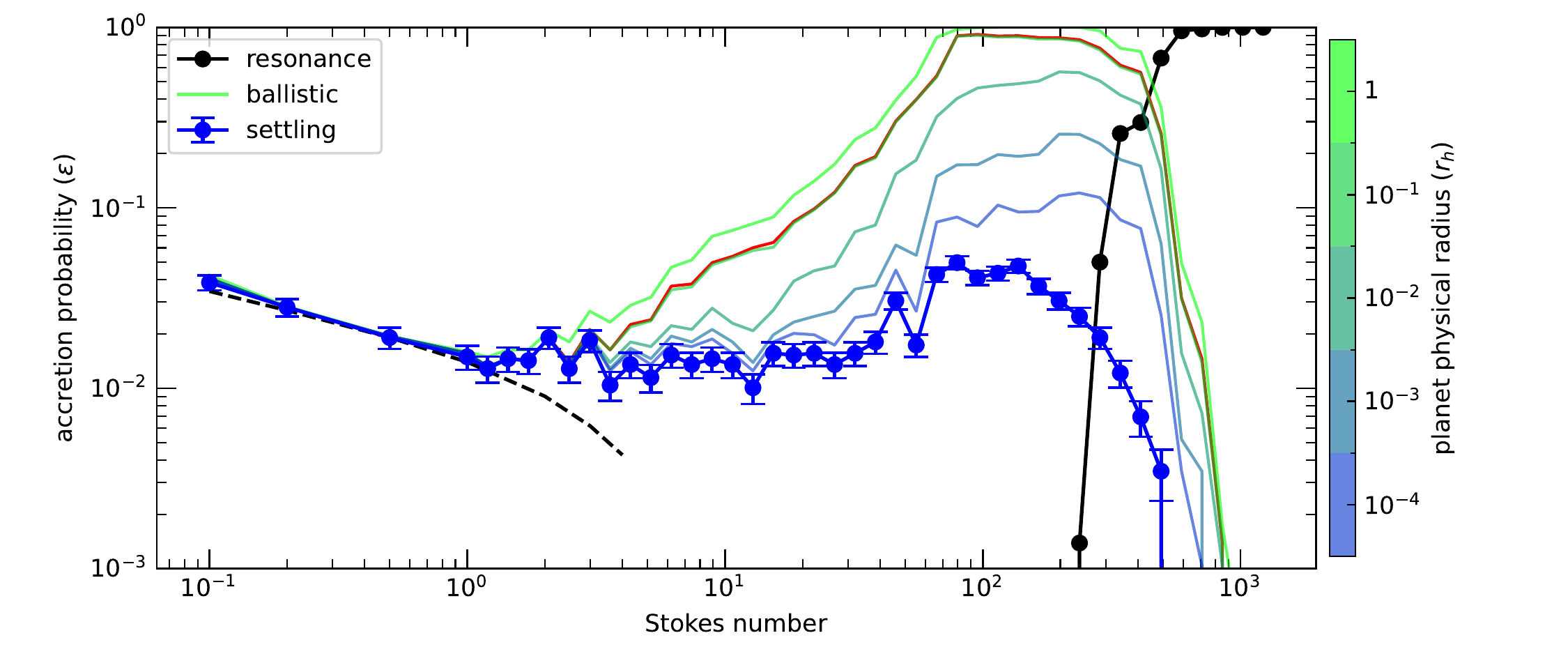}
        \caption{\label{fig:default} Pebble accretion efficiency \textit{vs} Stokes number in the default run. The parameters (i.e. $q$, $\eta$, $e_p$) used are listed in the first row of \Tb{parameters}. The $x$-axis is the Stokes number of drifting pebbles in the Epstein regime. The dashed black line for $\mathrm{St}<1$ represents the analytical expression by \citet{OrmelLiu2018}, valid for small pebbles.  
            Black curve, blue thick curve and thin curves denote probabilities of trapping into resonance, capture by settling and ballistic hits, respectively. The red thin line corresponds to a ballistic hit capture radius equal to the planet Bondi radius, while other thin lines correspond to physical radius of $(1,\ 0.1,\ 0.01,\ 10^{-3},\ 10^{-4}) r_\mathrm{h}$. 
        }
\end{figure*}

\fg{default} is the result for the default parameters: an Earth-mass planet on a circular orbit around a solar mass star, with pressure gradient parameter $\eta=3\times 10^{-3}$. We take the Epstein drag regime in this simulation. We also overplot the \citet{OrmelLiu2018} analytical prediction (black dashed line on the left) for the accretion rate, which fits the accretion efficiency of $\mathrm{St}<1$ well.  However, the behaviour at high Stokes numbers deviates from the low $\mathrm{St}$ case. Settling outcomes, when pebbles fall on to the planet regardless of its physical size, decrease in importance for larger St. Instead, ballistic hits takes over for $\mathrm{St}\gtrsim10$.  With increasing Stokes number, we can roughly divide the large pebble accretion into three stages: (i) `Ballistic rise' for $\mathrm{St}<70$, (ii) an `accretion plateau' when $70<\mathrm{St}<400$, and (iii) `resonance' when $\mathrm{St}>400$.

\underline{Ballistic rise}: Initially, when $\mathrm{St}<70$, pebbles still drift quickly and only a fraction of them will enter the Hill radius. The settling probability (blue dots) does not change much. This can be understood from the balance between the drift velocity and energy dissipation within the Hill sphere. As the drift velocity scales with $\mathrm{St}^{-1}$ when $\mathrm{St}>1$, the probability of entering the Hill sphere scales with $\mathrm{St}$. However, as the gas drag force scales with $\mathrm{St}^{-1}$ [\Eq{drag}], the energy dissipation rate inside Hill sphere,
\begin{equation}
    \dot{E} = F_\mathrm{drag}\Delta v \propto \mathrm{St}^{-1}
    \label{eq:dissipation}
\end{equation}
is inversely proportional to Stokes number. Assuming that the Jacobian energy is distributed uniformly, the settling probability is proportional to the energy dissipation rate.  Combining the above two effects, the settling probability is independent of Stokes number. 
On the other hand, the ballistic hit probability inside the Hill sphere does not rely on energy dissipation for large Stokes numbers. Therefore, the ballistic hit efficiency will scale with the probability of entering the Hill sphere, which increases with Stokes number. We indeed find that the measured ballistic $\epsilon$ are linear with St. 

\underline{Plateau stage}: for $70<\mathrm{St}<400$, the ballistic hit efficiency plateaus at about $50$ to $90$ per cent. In this stage, almost all pebbles enter the Hill radius because of their weak radial drift. The ballistic hit probability does not change much with Stokes number, resembling a plateau. This suggests that large pebbles with Stokes number in a range $70<\mathrm{St}<400$ can supply planet growth quite efficiently. Though settling is not comparably important in this stage, its probability still decreases with Stokes number, because of the lower energy dissipating rate for larger Stokes numbers.

\underline{Resonance stage}: when $\mathrm{St}>400$, both ballistic and settling efficiencies decrease, because pebbles get trapped in resonance. 
As the gas drag becomes weaker, radial drift is finally balanced by gravitational scattering from the planet. The pebbles are blocked by the planet from drifting and accreting.

The order of resonance varies with Stokes number. The larger Stokes number, the less gravitational perturbation needed to maintain resonance. Therefore the pebble can be trapped into resonance orbits further away from the planet, i.e. orbits with higher j (See \Fg{order}). Further more, even for pebbles with the same Stokes numbers on the verge of resonance, the order of resonance trapping is not deterministic (\Fg{order}). This is because as any initial positional correlation of the pebbles will be lost quickly due to chaotic resonance trapping. For example, $\mathrm{St}=1000$ pebbles can be trapped into $j=11, 12, 13$ (mainly) or (with lower probability) even higher-$j$ mean motion resonances.

Pebbles with orbits in different order of resonance may collide at high velocity. We examine the relative velocity at the point where two trajectories in the corotating frame intersect. The relative velocity at which pebbles collide is shown in \Fg{rv}. About half of the collisions have relative velocity larger than $0.1\ v_\mathrm{k}$, corresponding to $3000\ \mathrm{m}\ \mathrm{s}^{-1}$ at $1\ \mathrm{au}$ or $550\ \mathrm{m}\ \mathrm{s}^{-1}$ at $30\ \mathrm{au}$. Probabilistic trapping results in high relative velocity when large pebbles collide, causing fragmentation to smaller sizes and, ultimately, escape \citep{WeidenschillingDavis1985} to end up at Stokes numbers corresponding to the `plateau region'. 
\begin{figure}
    \centering
    \includegraphics[width=0.5\textwidth]{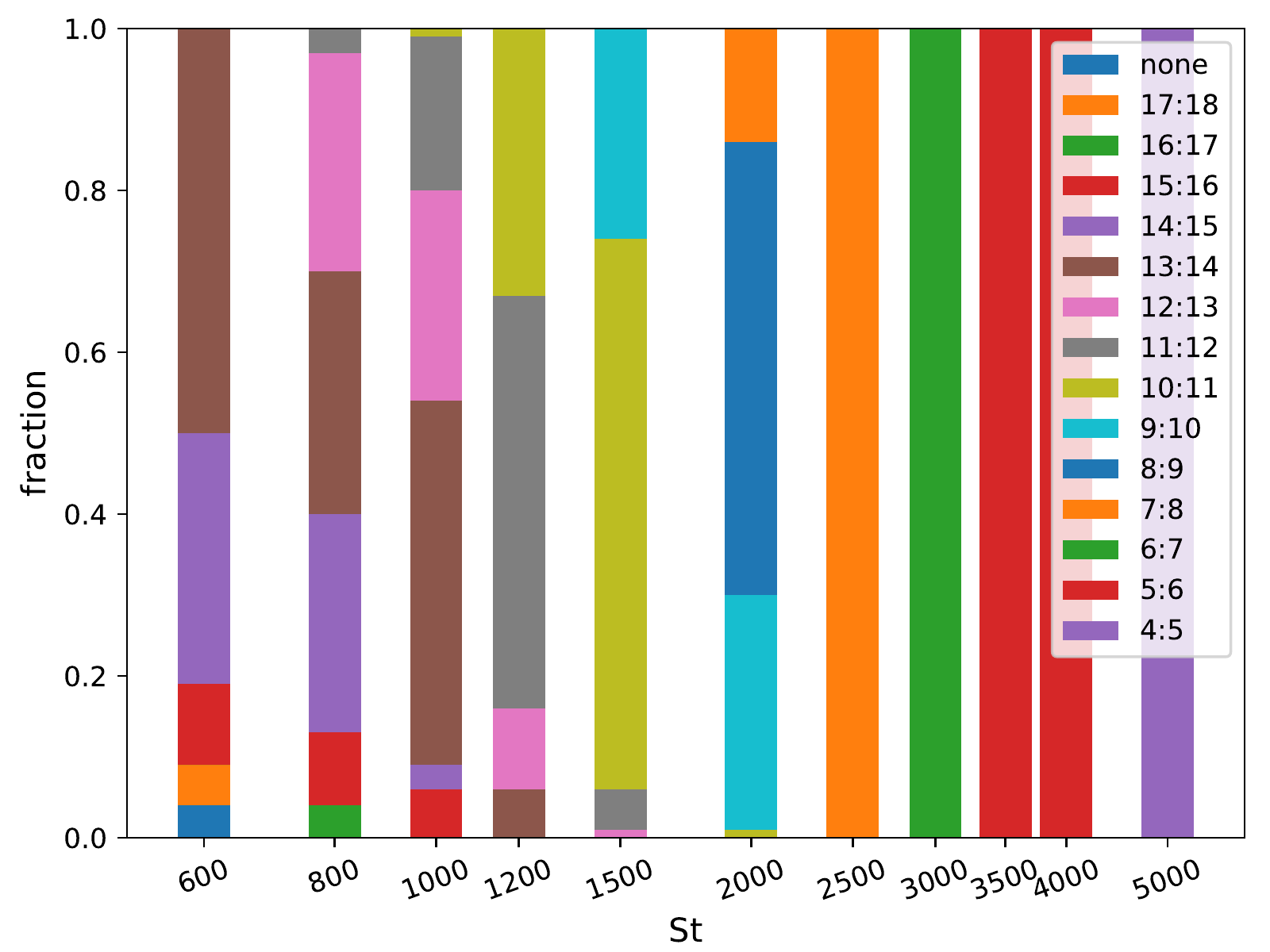}
    \caption{Order of mean motion resonance the pebbles are trapped in \textit{vs} Stokes number.}
    \label{fig:order}
\end{figure}
\begin{figure}
    \centering
    \includegraphics[width=0.5\textwidth]{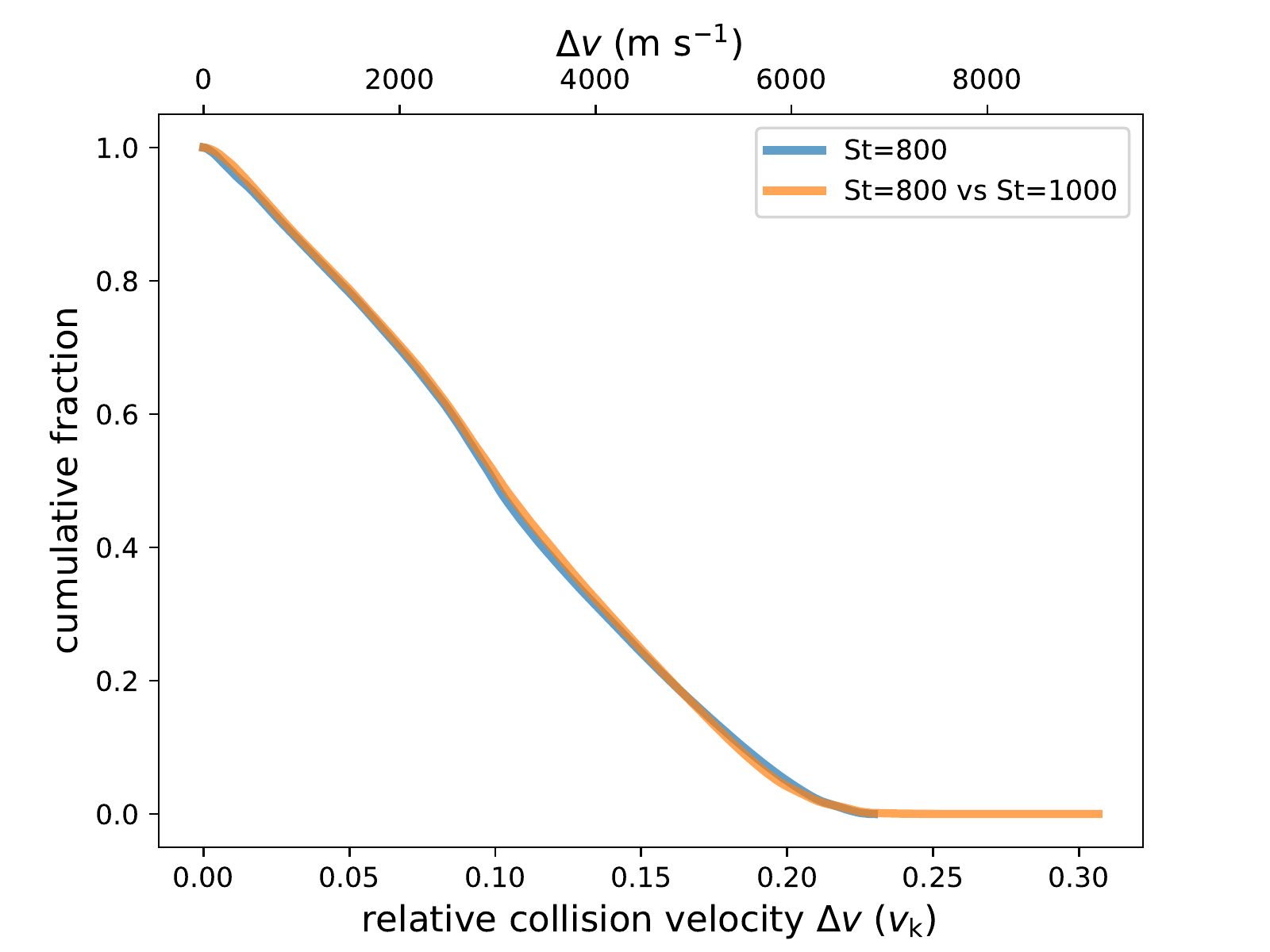}
    \caption{Cumulative distribution of the collision velocity of pebbles trapped in resonance. The $x$-axis shows the relative velocity $\Delta v$ at the collision in terms of the local Keplerian velocity and $\mathrm{m}\ \mathrm{s}^{-1}$. The y-axis gives the fraction of velocities exceeding $\Delta v$. Blue line shows the distribution among St=800 pebbles. The orange line shows the collisional velocity between St=800 pebbles and St=1000 pebbles.}
    \label{fig:rv}
\end{figure}

\subsection{Parameter study} \label{sec:parameter}
There are four free parameters besides $\mathrm{St}$ in our work: pressure gradient $\eta$, planet mass $q$, planet's orbital eccentricity $e_p$ and drag regime. In exploring each parameter, we fix other parameters at their default values. Note that as we used normalised units, planet orbital semi-major axis is not a free parameter here, though it may indirectly influence 
other parameters like $\eta$ and $\mathrm{St}$ [\eq{St}]. In this section, we only introduce our results for the Epstein drag law and describe the Stokes drag law results in Appendix \ref{sec:appendix}.

\subsubsection{Pressure gradient}
\begin{figure*}
    \centering
    \includegraphics[width=\textwidth]{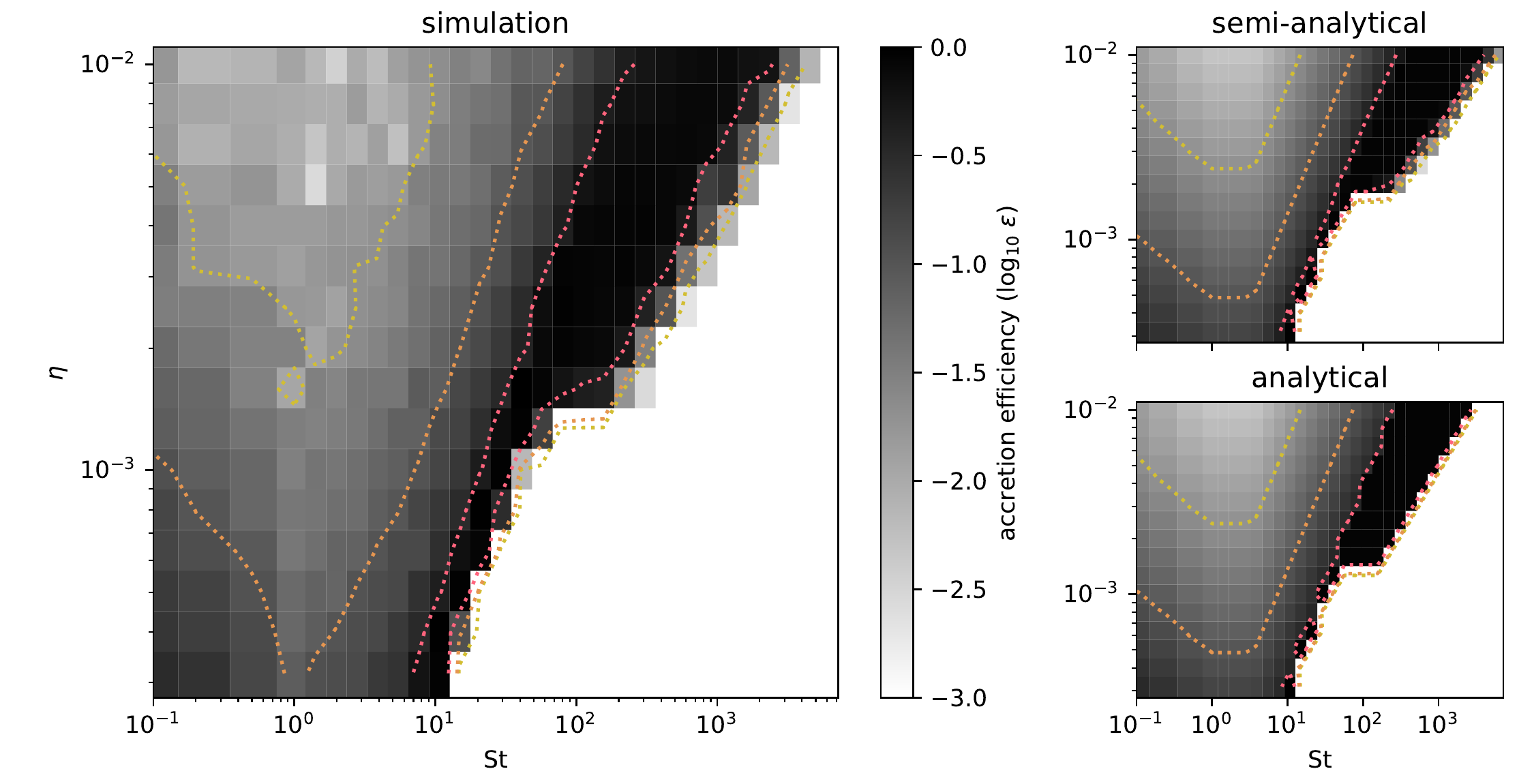}
    \caption{Accretion efficiency as function of pressure gradient parameter and Stokes number. The left panel is the simulation result. colours denote the accretion probability on a logarithmic scale, while yellow, orange and red contours show where the pebble accretion efficiency equals $0.02$, $0.1$ and $0.5$. Top right and bottom right panels are our semi-analytical model (\Se{semiana}) and the analytical fit (\Se{analytical}) for the accretion probability, respectively.}
    \label{fig:eta}
\end{figure*}
\Fg{eta} shows the results of varying the pressure gradient parameter $\eta$. We explored $\eta$ in the range of $10^{-3.5}$ to $0.01$. We find that with increasing $\eta$, the critical Stokes numbers where the accretion plateau and the resonance trapping start increase. This is because $v_\mathrm{drift}\propto \eta/\mathrm{St}$. When increasing $\eta$, $\mathrm{St}$ needs to increase correspondingly to keep the same drift speed.  The lowest Stokes number for resonance also increases with $\eta$ for the same reason. Yet at $\eta\sim 1.4\times 10^{-3}$ this interface suddenly shift to higher Stokes. This jump can be understood by comparing the eccentricity damping time-scale and the pebble's synodical time-scale. For $\eta\gtrsim 1.4\times 10^{-3}$ the resonance Stokes number is $\mathrm{St}_\mathrm{res}\gtrsim 200$, for which the eccentricity damping time-scale is longer than the synodical time-scale. The eccentricity associated with the resonance can be maintained, such that the mean motion resonance prevents pebbles from drifting inside. 
On the other hand, when the eccentricity is efficiently damped, no resonance is possible. The pebble's eccentricity is reduced to zero before being scattered by the planet \citep{MutoInutsuka2009}.  We also see that the accretion probability at the plateau region decreases with $\eta$. The higher drift velocities diminishes the probability of interaction with the planet. Still, the height of the accretion plateau is as high as $\epsilon=0.7$ even for $\eta = 7\times 10^{-3}$, demonstrating robust potential for planet growth with large pebbles. 

\subsubsection{Planet mass}
\begin{figure*}
    \centering
    \includegraphics[width=\textwidth]{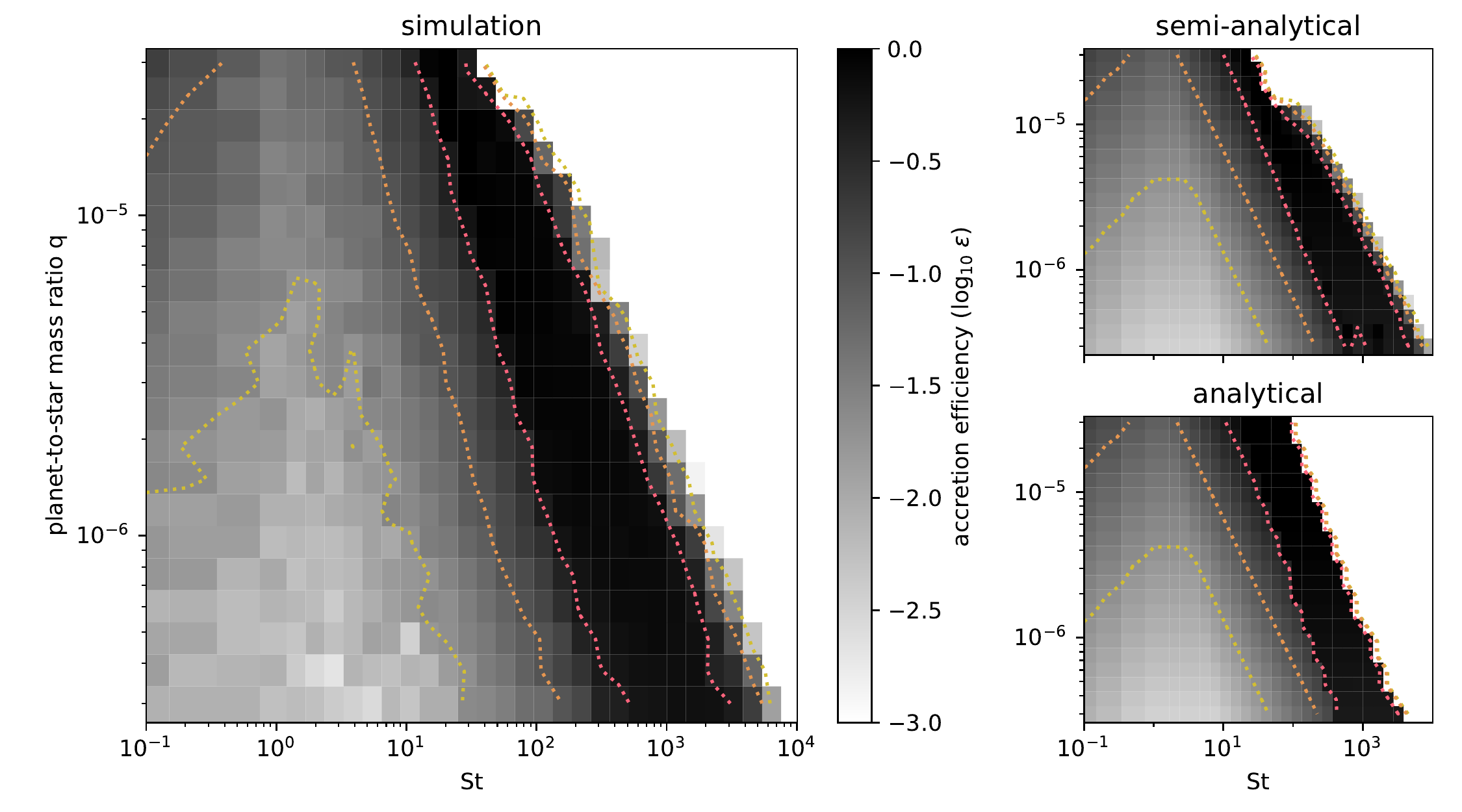}
    \caption{Same as \fg{eta}, but with the $y$-axis the planet-to-star mass ratio $q$.}
    \label{fig:q}
\end{figure*}

\Fg{q} shows the results of changing the planet-to-star mass ratio. Increasing the planet mass shifts the accretion plateau to smaller Stokes number. This is because the Hill radius increases as $q^{1/3}$. A more massive planet has a higher probability to capture the fast moving, smaller Stokes number pebbles. The Stokes number for the onset of resonance also decreases with planet mass because the heavier the planet the stronger the gravitational scattering of the pebbles, and the easier the pebble is trapped in resonance. Generally, if we increase the planet mass, the pebble accretion efficiency at the plateau stage increases because of the enhanced gravity. However, even for the lightest planet studied, $0.1M_\oplus$ ($q=3\times10^{-7}$), the ballistic hit probability still exceeds 50 per cent for Stokes numbers in the range $(500,\ 3000)$. Therefore throughout the planet growth process, from smaller mass to larger, the efficiency of large pebble accretion remains high.

\subsubsection{Eccentricity}
\begin{figure}
    \centering
    \includegraphics[width=0.5\textwidth]{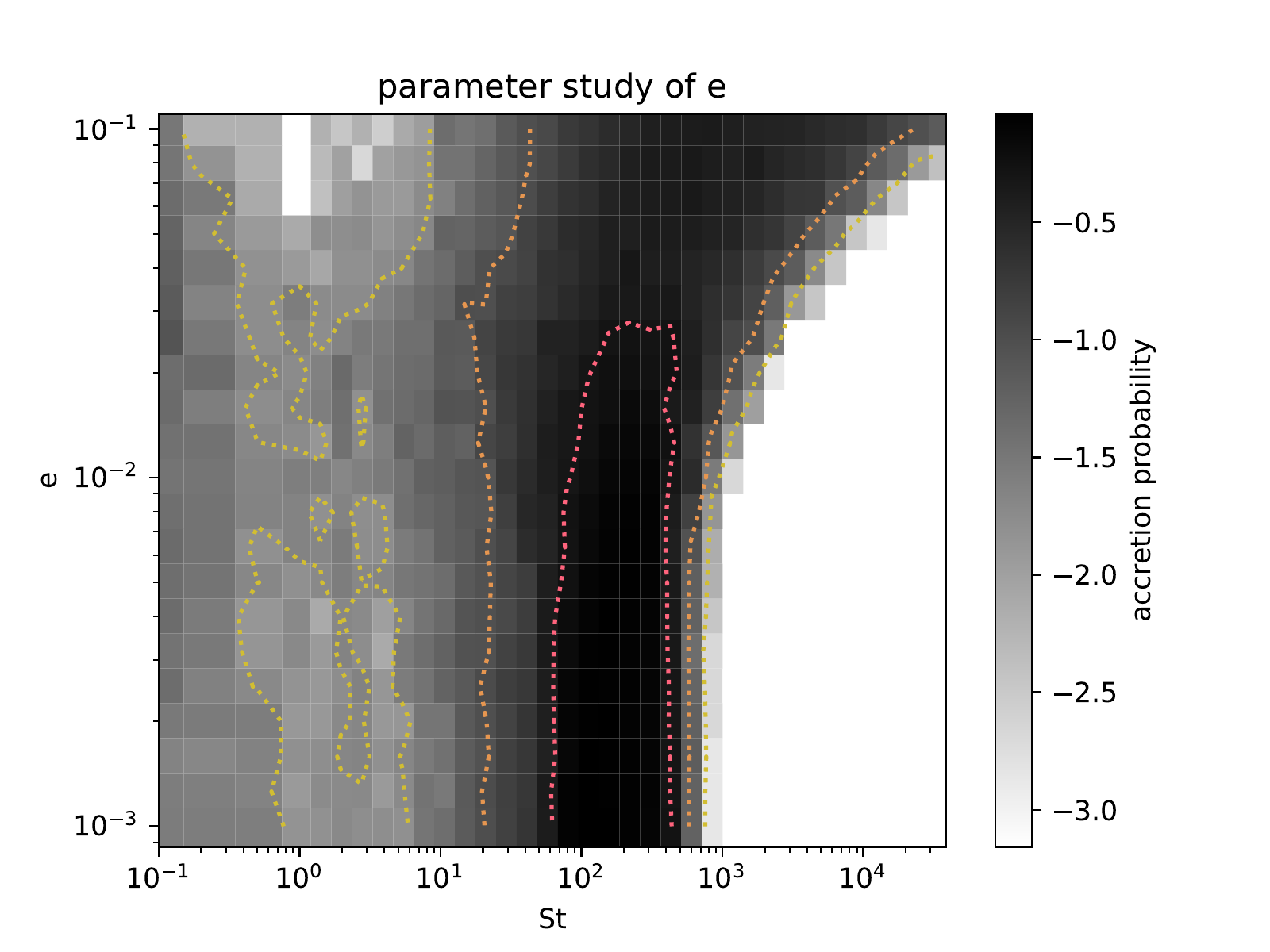}
    \caption{Same as the left panel of \fg{eta}, but with the $y$-axis the eccentricity of the planet $e_p$.}
    \label{fig:e}
\end{figure}

For planets on an eccentric orbit, we take the Bondi-Hoyle-Lyttleton radius as the physical radius of the planet, $r_\mathrm{BHL} = GM_p/(c_s^2+e^2v_\mathrm{k}^2)$. \fg{e} shows the results of changing the planet eccentricity. For eccentricities $e_p<0.01$, the simulation results converge to the circular planet case. This is because the relative velocity contribution from eccentricity, $e_pv_\mathrm{k}$, is negligible compared to the Hill velocity $r_\mathrm{h}\Omega$, which is the typical velocity through which pebbles enter the Hill sphere. As we increase the eccentricity to $0.03$, the eccentric relative velocity becomes increasingly important. The accretion plateau at $\mathrm{St}\sim 200$ decreases with planet eccentricity because the higher encountering velocity favours scattering over accretion. 
At the same time, the critical Stokes number at which pebbles get trapped in resonance increases, because the interactions with planets on eccentric orbits are more variable, rendering resonance trapping difficult. Specifically, if $er_p\gg r_\mathrm{h}$, the behaviour of pebbles becomes quantitatively different for the following reasons. 
First, the relative velocity becomes too large for settling which relies on dissipation of pebble's energy through gas drag. Second, the accretion probability at the plateau region decreases with eccentricity, because the planet size (Bondi-Hoyle-Lyttleton radius) decreases quadratically with eccentricity. Third, unlike circular planet case, where the resonant pebble is likely to have apsidal anti-aligned with the planet's, the pebbles trapped in resonance with eccentric planet are likely have their argument of pericentre aligned with the planet's, as \citet{LauneEtal2022} found. When $e_p=0.05$ and $\mathrm{St}=15000$, 53 per cent of the pebbles are trapped in resonances with apsidal alignment, among which most are in the first order. 
Fourth, apsidal antialigned pebbles are unlikely to be traped in the first order resonance. For the same $e_p$ and $\mathrm{St}$, around 13 per cent of all pebbles are trapped in second ($j{:}j+2$) and 25 per cent in third order ($j{:}j+3$) resonances, while few are trapped into first or order higher than four apsidal aligned resonances. 

\section{Analytical and semi-analytical model}

\subsection{Semi-analytical model for a planet on a circular orbit} \label{sec:semiana}

To estimate the accretion efficiency, we develop a semi-analytical model using the impulse approximation. A similar approximation is used by \citet{TanakaIda1997} to study the planetesimal distribution around a planet, but our model is tailored towards aerodynamically active pebbles. In our semi-analytical model, we separate the effect of gas drag and planet perturbation. When the pebble is far from the planet, we neglect the planet's gravity and only account for its drift motion.
On the other hand, when the pebble comes close to the planet, we only employ the gas-free orbital perturbation by the planet as for aerodynamically large pebbles the encounter time-scale is much shorter than the pebble stopping time.  Thus the motion of a pebble is divided into two stages: (i) radial drift due to drag over a synodical time, and (ii) an instantaneous kick at the conjunction.

In order to obtain the change in the orbital elements after each encounter, we employ the findings of \citet{HasegawaNakazawa1990}, neglecting gas drag during the encounter. A 2D trajectory of a pebble in a Keplerian orbit is characterised by three orbital elements: the semimajor axis and the two components of the eccentricity vector $\bm{e}=(e_1,e_2)$, where $\bm{e}$ is directed to the pericentre of the orbit. Writing the above three orbital elements in Hill units, $a \equiv  r_p+b r_\mathrm{h}$ and $e_i = p_i r'_\mathrm{h}\ (i=1,\ 2)$, \citet{HasegawaNakazawa1990} derived the change in $p_1$ and $p_2$ after each encounter with the planet under Hill's approximation and perturbation theory:
\begin{eqnarray}
    \begin{aligned}
    \Delta p_1 &= -\frac{9}{b^3}\left[\left(\frac{2}{9}-\frac{1}{2}R_2\right)p_2-R_4\frac{p_1p_2}{b} \right] \\
    \Delta p_2 &= -\frac{9}{b^2}\left[-R_1-\left(\frac{2}{9}+\frac{1}{2}R_2\right)\frac{p_1}{b}+R_6\frac{p_1^2}{b^2}+R_7\frac{p_2^2}{b^2}\right] ,
    \end{aligned}
    \label{eq:Hasegawa}
\end{eqnarray}
where the $R_i$'s are constants listed in the Appendix of \citet{HasegawaNakazawa1990}. Then, we calculate the change in semimajor axis from the conservation of the Jacobi integral,
\begin{equation}
    \Delta b^2 = \frac{4}{3}\Delta p^2,
    \label{eq:Hasegawa2}
\end{equation}
where $p^2=p_1^2+p_2^2$.

Away from the planet, the unperturbed radial and azimuthal equations of motion in the corotational frame are 
\begin{align}
    \frac{da}{dt} &= -\frac{2\eta \mathrm{St}}{\mathrm{St}^2+1}\left(\frac{a}{r_p}\right)^{-\frac{1}{2}}v_\mathrm{k} \label{eq:reom}\\
    \frac{d\theta}{dt} &= \left(\frac{a}{r_p}\right)^{-\frac{3}{2}}-1.
\end{align}
Solving for the polar angle $\theta$, we get
\begin{equation}
    \theta (t) = \theta_0-\Omega_\mathrm{k}t-\frac{\mathrm{St}^2+1}{3\eta \mathrm{St}} \ln{\left[1-\frac{3\eta \mathrm{St}}{\mathrm{St}^2+1}\frac{\Omega_\mathrm{k}t}{\left(b_0r'_\mathrm{h}+1\right)^{\frac{3}{2}}}\right]}
    \label{eq:ttheta}
\end{equation}
where $b_0$, $\theta_0$ are the normalised semi-major axis and polar angle of the pebble in the corotational frame at $t=0$. Between two conjunctions $\theta (t)$ changes from $2\pi$ to $0$. 
Thus we can solve for the time interval between two conjunctions $t_\mathrm{syn}$ with
\begin{equation}
    \theta(t_0+ t_\mathrm{syn}) - \theta(t_0) =-2\pi.
    \label{eq:tinterval}
\end{equation}

We then solve \Eq{reom} to calculate the pebble's semimajor axis after a synodical time. A special situation arises when the pebble drifts too fast, such that it ends up interior to the planet's orbit. In that case, 
$\theta(t)$ reaches a minimum at $\Omega_{k}t_\mathrm{min}=[(b_0r'_\mathrm{h}+1)^\frac{3}{2}-1](\mathrm{St}^2+1)/(3\eta \mathrm{St})$. Hence, when $\theta(t_\mathrm{min})>\theta_0-2\pi$, \Eq{tinterval} does not have a solution. 

In order to calculate the eccentricity damping in the linear drag regime, we employ the eccentricity damping time-scale $t_e=\mathrm{St}/\Omega_\mathrm{k}$ \citep{Adachi1976}. Thus the eccentricity after a synodical time $t_\mathrm{syn}$ is reduced by $\mathrm{e}^{-t_\mathrm{syn}/t_e}$. We repeat this procedure of drift-and-kick to obtain the pebble's orbital elements just before and after each conjunction. 

The semi-analytical model above is only a model for the approaching and scattering, but not for the capture or ballistic hit inside the Hill radius. This is because \Eq{Hasegawa} approximates planet gravity as perturbation to the pebble's heliocentric orbital elements, $b$ and $\bm{p}$, which is not the case when planet's gravity dominates inside the Hill radius. 
Therefore, to derive an expression for accretion rate in ballistic rise stage, we resort to extending the capture and ballistic rate at $\mathrm{St}=1$, and utilise the semi-analytical model to correct the results.

If the planet's perturbation is neglected, the ballistic hit efficiency is proportional to $v_\mathrm{drift}^{-1} \propto \mathrm{St}$. Thus the accretion probability in the ballistic rise stage can be written as 
\begin{equation}
    \epsilon_\mathrm{rise}^\mathrm{ana} = \max \{\epsilon_\mathrm{set,1},\ \mathrm{St}\ \epsilon_\mathrm{bal,1}\}
    \label{eq:rise}
\end{equation}
where $\epsilon_\mathrm{set,1}$ and $\epsilon_\mathrm{bal,1}$ are the 2D settling and ballistic hit probability of $\mathrm{St}=1$ pebble from \citet{LiuOrmel2018}:
\begin{equation}
    \begin{aligned}
    \epsilon_\mathrm{set} &= 0.32\sqrt{\frac{q\Delta v}{\mathrm{St}\eta^2 v_\mathrm{k}}}\\
    \epsilon_\mathrm{bal} &= \frac{R_p/r_p}{2\pi\eta\mathrm{St}} \sqrt{2qr_p/R_p+\Delta v^2 /v_\mathrm{k}^2}.
    \end{aligned}
\end{equation}
Here $R_p$ is the planet physical radius and $\Delta v$ is the encounter velocity between planet and pebble, consisting of a headwind and a shear component:
\begin{equation}
    \frac{\Delta v}{v_\mathrm{k}} = \frac{\eta}{1+5.7q\mathrm{St}/\eta^3} + 0.52(q\mathrm{St})^{\frac{1}{3}}.
\end{equation}

Then we correct \eq{rise} by the fractions of `strong kick' events, obtained from the semi-analytical model. Gravitational scattering from the planet may `kick' the pebble outward and decrease its net drift velocity. This scattering can be so strong that the pebble is `kicked' to such a high orbit at which it has another chance to enter the Hill radius.  We define `strong kick' events when the kick by the planet is larger than its radial drift during the subsequent synodical interval. If the strong kick occurrence rate is denoted $f_\mathrm{strong}$, $\epsilon_\mathrm{rise}^\mathrm{ana}$ is corrected by a factor $f_\mathrm{per}^\mathrm{semi} = 1+f_\mathrm{strong}$: 
\begin{equation}
    \epsilon_\mathrm{rise}^\mathrm{semi} = \epsilon_\mathrm{rise}^\mathrm{ana}(1+f_\mathrm{strong}),
\end{equation}
In the semi-analytical model, the fractions $f_\mathrm{strong}$ follow 
\begin{equation}
        f_\mathrm{strong} = N_\mathrm{strong}/N,
\end{equation}
where $N_\mathrm{strong}$ is the total count of `strong kick' events and $N$ is the number of pebbles calculated with the semi-analytical model.
For example, $f_\mathrm{strong}$ increases from 0.056 to 0.64 when $\mathrm{St}$ increase from 40 to 80, because slower drifting pebbles have higher chance to interact with the planet.

The accretion probability at the plateau region is calculated by our analytical fit given in the next section [\eq{analytical}]. Beyond the plateau, resonance trapping reduces the accretion efficiency by a factor $1-\epsilon^\mathrm{semi}_\mathrm{res}$, where $\epsilon^\mathrm{semi}_\mathrm{res}$ is the probability to be captured in resonance, which we determine semi-analytically. The criterion of resonance is adopted identical to that in the simulation: the measured mean drift velocity falls below $10^{-4}v'_r$. However, the drift-and-kick model is only valid when the pebble is not too close to the planet \citep{HasegawaNakazawa1990}. For close encounter, even though the pebble enters the Hill sphere, breaking the resonance in the full simulation, we still assume the pebble is outside the Hill sphere and calculate the kick in semi-analytical model. 
Thus to calculate resonance fraction, we further require the impact parameter between pebble and the planet before each conjunction larger than $f r_\mathrm{h}$, with $f=1.65$ empirically. 

We run 2000 realisations for $\mathrm{St}=10$ pebbles with default parameter to get semi analytical accretion rate, which takes 3.3 seconds on a loptop PC, compared to 109 seconds with the full integration. The results of our semi-analytical prediction are shown in the top right panels of \fgs{eta}{q}. For each horizontal row [fixed $(q,\ \eta)$ but varying $\mathrm{St}$], the ballistic rise, plateau (dark region between $\epsilon=0.5$ contours), and resonance stages are demonstrated by the semi-analytical model.  The trend that the accretion probability increases with $q$ and decreases with $\eta$ before resonance is also reproduced well by the semi-analytical model.  Most of the predictions agree with the numerical result within a factor of 1.5. The semi-analytical model can also reproduce the resonance Stokes number, including the discontinuity caused by the different eccentricity damping regimes at $\eta\sim 1.4\times 10^{-3}$. 

\subsection{Analytical fit to the results} \label{sec:analytical}
\begin{table}
    \caption{Notation used in analytical fit}
    \label{tab:analytical}
    \centering
    \small
    \begin{tabular}{l l p{5.5cm} }
        \hline
        label & meaning\\
        \hline
        $p_\mathrm{s}$ & Eq.21 &settling probability after the pebble enters Hill sphere\\
        $p_\mathrm{b}$ & Eq.21 &ballistic hit probability after the pebble enters Hill sphere\\
        $p_\mathrm{i}$ & Eq.21 &probability to be scattered inward\\
        $R_p$ & & planet physical radius\\
        $A_\mathrm{p}$ & $5.56$     & fit constant\\
        $A_\mathrm{s}$ & $3.8$      & fit constant\\
        $A_\mathrm{b}$ & $2.36$     & fit constant\\
        $A_\mathrm{i}$ & $0.610$    & fit constant\\
        $A_\mathrm{wd}$ & $3.5$     & fit constant\\
        $A_\mathrm{sd}$ & $0.33$    & fit constant\\
        \hline
    \end{tabular}
\end{table}

\begin{table}
    \caption{Summary of the analytical fit for the accretion efficiency}
    \label{tab:ana_sum}
    \centering
    \small
    \begin{tabular}{p{5.2cm} p{2.4cm} }
        \hline
        1. Find the boundary between rising and plateau stage. $\mathrm{St}_\mathrm{pla}$ & \eq{St_pla}\\
        2. Find the boundary between plateau and resonant stage in weak damping regime $\mathrm{St}_\mathrm{res,weak}$. & Eqs. (\ref{eq:Hasegawa}, \ref{eq:Hasegawa2}, \ref{eq:St-res-weak}, \ref{eq:St-res-weak-ini}) \\
        3. If weak damping criterion \eq{St-crit} not satisfied, calculate $\mathrm{St}_\mathrm{res,strong}$. & \eq{St-res-str} \\
        4. Accretion efficiency in rising stage $\epsilon_\mathrm{rise}^\mathrm{ana}$. & \eq{rise} \\
        5. Accretion efficiency in plateau stage $\epsilon_\mathrm{rise}^\mathrm{pla}$.& \eq{analytical}\\
        \hline
    \end{tabular}
\end{table}

With further approximations, we can write down a fully analytical expression for the accretion probability. We summarize our analytical fit in \Tb{ana_sum}\footnote{A python script implementation is available at \url{https://github.com/chrisormel/astroscripts/tree/main/papers}.}. For the accretion plateau, the drifting velocity of the pebble is so small that it will always enter the Hill sphere. Once it enters the Hill sphere, the probability to be captured through settling $p_\mathrm{s}$ is proportional to $\mathrm{St}^{-1}$ [\eq{dissipation}]. If the pebble is not accreted, it is scattered either inward or outward, at probability $p_\mathrm{i}$ and $1-p_\mathrm{i}$, respectively. 
Once it is scattered inward, it will drift to the star, with no chance to be accreted again, unless the planet is on an eccentric orbit. On the other hand, a pebble scattered outward will have another chance to enter the Hill radius.  During each scatter event, the pebble has a probability ${\propto}\sqrt{R_p}$ to ballistically hit the planet, following the gas-free 2-body gravitational interaction assumption \citep{Safronov1972}
Thus, the accretion probability for the plateau region becomes a summation of an infinite series:
\begin{equation}
    \displaystyle
    \epsilon^\mathrm{ana}_\mathrm{plat} = 1 - \frac{(1-p_\mathrm{s})(1-p_\mathrm{b})p_\mathrm{i}}{1-(1-p_\mathrm{s})(1-p_\mathrm{b})(1-p_\mathrm{i})}
    \label{eq:analytical}
\end{equation}
where
\begin{equation}
    \begin{aligned}
        p_\mathrm{s} &= A_\mathrm{s} \mathrm{St}^{-1}\\
        p_\mathrm{b} &= A_\mathrm{b}\left(\frac{R_p}{r_\mathrm{h}}\right)^{1/2}\\
        p_\mathrm{i} &= A_\mathrm{i}
    \end{aligned}
\end{equation}
The meanings of notations are listed in \Tb{analytical}.

Next, we analytically fit the lower and upper boundaries of the accretion plateau, $\mathrm{St}_\mathrm{plat}$ and $\mathrm{St}_\mathrm{res}$. The criterion for the lower boundary is that every pebble will drift into the Hill sphere, i.e. $t_\mathrm{syn}v_\mathrm{dr} \sim r_\mathrm{h}$. As the drift velocity is $v_\mathrm{dr}\sim 2\eta v_\mathrm{k}/\mathrm{St}$ and the synodical time-scale is $t_\mathrm{syn}\sim 4\pi/3\ (r_\mathrm{h}/r_p)^{-1}\ \Omega_\mathrm{k}^{-1}$, it follows that the boundary Stokes number is
\begin{equation}
        \mathrm{St}_\mathrm{plat} = A_\mathrm{p}\frac{\eta}{q^{2/3}}, 
        \label{eq:St_pla}
\end{equation} 
where the proprtional constant $A_\mathrm{p}$ is fit to be 5.56.

For the high-Stokes boundary of the plateau region ($\mathrm{St}_\mathrm{res}$) we distinguish the weak and strong eccentricity damping regimes. In the strong damping limit when the Stokes number is smaller than a critical number $\mathrm{St}_\mathrm{crit}$, the pebble's orbit is circularised within succesive conjunctions. Therefore before each encounter, its eccentricity is neglected. On the other hand, in the weak damping limit, the pebble can sustain its eccentricity before getting another kick.

The $\mathrm{St}_\mathrm{crit}$ is determined by comparing the eccentricity damping time-scale to the synodical time-scale, $t_e = t_\mathrm{syn}$. We take the impact parameter to be $b_\mathrm{crit} = 2.5$ \citep{IdaNakazawa1989} and thereby calculate the synodical time-scale $t_\mathrm{syn}$, hence, 
\begin{equation}
    \label{eq:St-crit}
    \mathrm{St}_\mathrm{crit} = \frac{4\pi}{3}\frac{r_\mathrm{p}}{2.5r_\mathrm{h}} = 168\left(\frac{q}{3\times 10^{-6}}\right)^{-\frac{1}{3}}.
\end{equation}

If eccentricity damping is weak, we equate the semi-major axis change after scattering [\Eqs{Hasegawa}{Hasegawa2}] with the radial drift within one synodical time-scale,
\begin{equation}
    \Delta b_\mathrm{kick} = \frac{v_\mathrm{r}t_\mathrm{syn}}{r_\mathrm{h}} = \frac{8\pi\eta}{3b\mathrm{St}{r'_\mathrm{h}}^2},
\end{equation} 
to solve for the critical Stokes number in the weak damping regime $\mathrm{St}_\mathrm{res,weak}$, 
\begin{equation}
    \label{eq:St-res-weak}
    \mathrm{St}_\mathrm{res,weak} = A_\mathrm{wd}\frac{8\pi \eta}{3b \Delta b_\mathrm{kick}{r'_\mathrm{h}}^2}.
\end{equation}
When using \Eq{Hasegawa2} for $\Delta b_\mathrm{kick}$, we take the eccentricity of the pebble to be the equilibrium eccentricity corresponding to the j:j+1 resonance \citep{GoldreichSchlichting2014, TerquemPapaloizou2019, HuangOrmel2023}, $e_\mathrm{eq} = \sqrt{t_e/(2jt_a)}$,
where $\Omega_\mathrm{k}t_e = \mathrm{St}$ and $\Omega_\mathrm{k}t_a = \mathrm{St}/(2\eta)$ are the eccentricity and semi-major axis damping time-scales. Using Kepler's third law $(j+1)/j=(1+br'_\mathrm{h})^{3/2}$, this gives
\begin{equation}
    e_\mathrm{eq} = \sqrt{\frac{3\eta b r'_\mathrm{h}}{2}}.
\end{equation}
We take normalised encounter semi-major axis to be $b=1.9$, which differs from $b_\mathrm{crit} = 2.5$ because in the weak damping regime the pebble has a non-negligible eccentricity \citep{IdaNakazawa1989}.  At the verge of resonance, the resonance angle -- or the mean anomaly at conjunction -- is $\pi/2$ \citep{TerquemPapaloizou2019, HuangOrmel2022}. Therefore, we approximate the eccentricity component in \Eq{Hasegawa} as 
\begin{equation}
    \label{eq:St-res-weak-ini}
    \begin{aligned}
        p_1 &= 0\\
        p_2 &= \sqrt{\frac{3\eta b}{2r'_\mathrm{h}}}.
    \end{aligned}
\end{equation}
Inserting these expressions in \eq{Hasegawa} we obtain the change in the semi-major axis $\Delta b_\mathrm{kick}$. Then, \eq{St-res-weak} with the empirically fitted correction factor $A_\mathrm{wd}=3.5$ provides the boundary between the plateau and resonance regimes in the weak damping limit.
When the resonance Stokes number in the weak damping limit is sub-critical, i.e. $\mathrm{St}_\mathrm{res,weak}<\mathrm{St}_\mathrm{crit}$, weak damping assumption no longer holds. Conversely, for the strong damping case, the eccentricity before cunjunction is close to zero. We therefore can adopt the averaged orbital element change calculated by \citet{MutoInutsuka2009}, which, in the large Stokes number limit, reads
\begin{equation}
    \left<\frac{da}{dt}\right> = -\frac{2\eta}{\mathrm{St}}v_\mathrm{k} + \frac{\alpha}{t_\mathrm{syn}}\frac{r'_\mathrm{h}}{b^5},
\end{equation}
where $\alpha\approx 30$. The radial motion of the gas as well as the planet's spiral density wave are neglected. 
Setting $\left<\frac{da}{dt}\right>=0$, we get
\begin{equation}
    \label{eq:St-res-str}
    \mathrm{St}_\mathrm{res, strong} = A_\mathrm{sd} \frac{8\pi}{3\alpha} \eta\frac{b_\mathrm{crit}^4}{{r'_\mathrm{h}}^2}.
    = 36.0\left(\frac{\eta}{10^{-3}}\right)\left(\frac{q}{3\times 10^{-6}}\right)^{-\frac{2}{3}}
\end{equation}
The proportionality constant $A_\mathrm{sd}$ is fit to be $0.33$. 

For $\mathrm{St}<\mathrm{St}_\mathrm{plat}$, we take \Eq{rise}, while for $\mathrm{St}>\mathrm{St}_\mathrm{res}$, the accretion probability is set to 0.  

The analytical prediction results are shown in the bottom right panels of \fgs{eta}{q}.  For most grid points the analytical fit expressions fall within 50 per cent of the simulation results. The analytical expression fit the plateau and rising stages well, except near the plateau boundary where the analytical model underestimates the accretion efficiency because we do not consider the planet perturbation to the pebble orbit, which decreases the net drift velocity, as we discussed in semi-analytical model. The resonant Stokes number $\mathrm{St}_\mathrm{res}$ is well reproduced analytically, supporting the point that resonance cannot be established if the eccentricity is damped within a synodical time-scale. 
\begin{figure}
    \centering
    \includegraphics[width=\columnwidth]{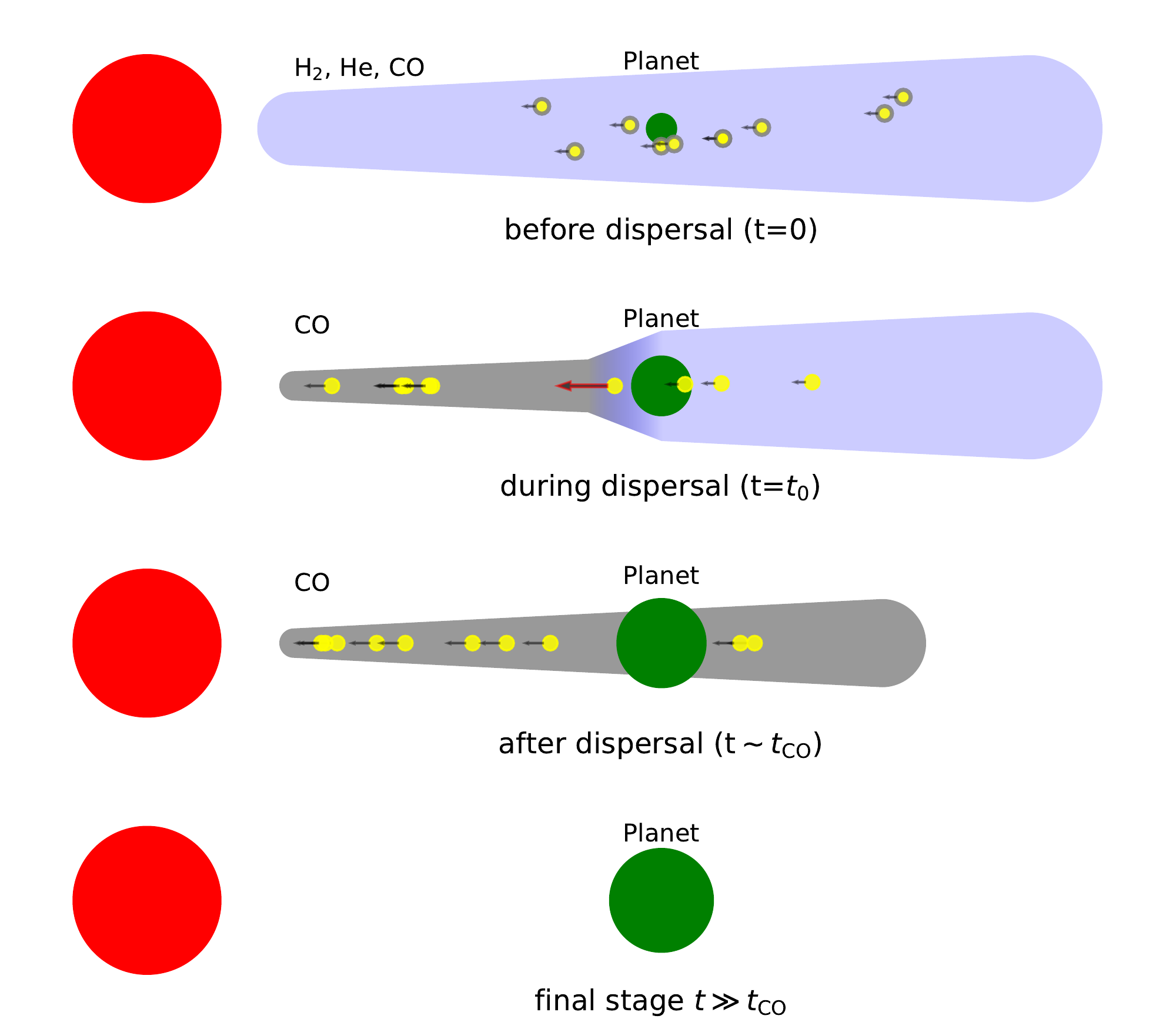}
    \caption{\label{fig:application-setup}Schematic of the late accretion process. Red, green, and yellow circles denote, respectively, the star, planet and small dust particles. During photoevaporation, the dust particles quickly increase their Stokes number to drift inwards, for pebble accretion to transition to the $\mathrm{St}>1$ mode.  We investigate how much of the planet final mass is contributed by large pebbles.}
\end{figure}

\section{Application to late accretion of small particles} \label{sec:application}
In this section, we apply our findings on the accretion of aerodynamically large pebbles to the late stage of planet formation, where we envision a remnant population of ${\sim}10\,\mu\mathrm{m}$ dust particles to become dynamically active. Usually, pebble accretion is thought to be driven by particles that grew in the outer regions of protoplanetary discs, before they drift inwards to be accreted by planets. 
However, small dust particles may have Stokes numbers too small to drift significantly during the protoplanetary disc phase.  They therefore remain in the outer disc until the gas dissipates, after which they transition to large Stokes number pebbles ($\mathrm{St}>1$) for which they could be accreted efficiently.  We focus on this leftover population of small dust in the disc and investigate their accretion by a planet, i.e., how they contribute to the planet's growth as aerodynamically large pebbles.

Specifically, we hypothesize a scenario in which the outer disc is characterised by particles that are small (radius $a\sim$30\,$\mu$m) such that they experienced minimal drift during the primordial, H/He-dominated, gas disc phase, but became aerodynamically active upon disc dispersal.  We consider how these dust particles are accreted by a planet with an initial mass $M_{p0}$ at 20\,au. We identify four stages as shown in \fg{application-setup}. (i) Initially, while the primordial gas disc is still around, accretion operates in the small Stokes limit. 
(ii) During the dispersal phase, the primordial gas in the disc photoevaporates, leaving a secondary disc replenished by outgassing. The Stokes number increases rapidly (the size remains fixed in our model), reaching, and then exceeding, $\mathrm{St}=1$. 
(iii) Afterwards, in the debris disc, the dust accretion operates in the large Stokes regime. (iv) Gas in the debris disc finally disperses, halting the drift of dust as they reach $\mathrm{St}\gg 1$.  

\subsection{Gas model}\label{sec:application-gas}
The fraction of transition disc given by SED observations indicates that most discs disperse on a time-scale on the order of ${\sim}10^5\,\mathrm{yr}$ \citep{ErcolanoPascucci2017}.  One mechanism to transition from protoplanetary discs to debris discs is photoevaporation. During photoevaporation FUV, EUV and X-ray photons from both the central star and the external stellar environment dissipate primordial H, He in the disc\citep{CarreraEtal2017,ColemanHaworth2022,AlexanderEtal2006i,GortiEtal2015}.
According to photoevaporation theory a gap will open in the disc at several au, when the disc accretion flux decreases below the photoevaporation mass loss rate. Material interior to it is drained quickly by viscous processes while the more loosely bound primordial gas of the outer disc is blown away by UV radiation \citep{OwenEtal2010}. The disc has turned into a debris disc, with its mass dominated by solids. 
\citet{MichelEtal2021} found that the average mass of dust particles with mm-size for Class-III discs is $0.29\ M_\oplus$. \citet{HalesEtal2022} fitted the ALMA continuum of HD 110058 debris disc, and find the mass of particles up to cm-size to be $0.08\ M_\oplus$.

\begin{table}
    \caption{Default parameters used in application}
    \label{tab:application parameters}
    \centering
    \small
    \begin{tabular}{c r p{4.5cm}}
    \hline
    label & value & description\\
    \hline
    $M_\ast$ & $M_\odot$ & Stellar mass\\
    $M_\mathrm{disc}$ & $0.1M_\odot$ & Disc total mass\\
    $R_1$ & $40\ \mathrm{au}$ & Disc characteristic radius scale\\
    $r_\mathrm{in}$ & $10\ \mathrm{au}$ & Inner boundary of the simulation\\
    $r_\mathrm{out}$ & $1000\ \mathrm{au}$ & Outer boundary of the simulation\\
    $\alpha$ & $0.001$ & viscous parameter\\
    $\alpha_z$ & $0.001$ & vertical turbulent parameter\\
    $t_0$ & $2\ \mathrm{Myr}$ & Time when photoevaporation front pass the planet\\
    $t_\mathrm{pe}$ & $10^4\ \mathrm{yr}$ & Hydrogen helium depletion time-scale\\
    $r_{\mathrm{f},i}$ & $3\ \mathrm{au}$ & Initial position of photoevaporation gap\\
    $\dot{M}_{\mathrm{pe},i}$ & $6.75\times10^{-9}\mathrm{M}_\odot\ \mathrm{yr}^{-1}$ & Photoevaporation mass loss rate\\
    $\rho_\cdot$ & $1\ \mathrm{g}\ \mathrm{cm}^{-1}$ & internal density of the particle\\ 
    $f_\mathrm{dg}$ & $0.0003$ & Dust-to-gas ratio of small particles\\
    $M_{p0}$ & $1M_\oplus$ & Planet initial mass\\
    $f_\mathrm{CO}$ & $0.1$ & CO ice fraction in the dust particles\\
    $a$ & $30\ \mu\mathrm{m}$ & particle size\\
    $t_\mathrm{CO}$ & $2\times 10^6\ \mathrm{yr}$ & CO lifetime in the debris disc\\
    \hline
    \end{tabular}
\end{table}

Yet debris discs may not be devoid of gas, as recent observations found evidence of tracer gas species, e.g., $\mathrm{CO} (\mathrm{J}=2-1)$ \citep{k_sp_l_2013,MoorEtal2017}, $\mathrm{CO} (\mathrm{J}=3-2)$ \citep{DentEtal2014,HughesEtal2017,HiguchiEtal2019ii}, and the $^3\mathrm{P}_1$ -- $^1\mathrm{P}_1$ transition line of $^{13}\mathrm{C}$ \citep{HiguchiEtal2019i}. Gas-rich debris discs like HD 141569A and HD 110058 are found to host CO gas with mass on the order of $0.1\ M_\oplus$ \citep{DiFolcoEtal2020, HalesEtal2022}.
Whether gas in debris disc is the remnant of the protoplanetary disc \citep{NakataniEtal2021} or has a secondary origin, like outgassing from dust or destruction of planetesimals \citep{MarinoEtal2016, Wyatt2020}, remains an open question. If its origin is primordial, CO is shielded from the stellar irradiation by $\mathrm{H}_2$, carbon grains or other CO molecules \citep{VisserEtal2009}. In the second senario, CO is released when volatile-rich solids in the debris disc grind down through collisions \citep{KralEtal2016}.  Regardless its origin, the debris disc gas may allow radial drift and continual accretion of pebbles after the protoplanetary disc phase. 

For the gas profile before dispersal, we take the self-similar solution of viscous evolution from \citet{Lynden-BellPringle1974}:
\begin{equation}
    \Sigma_v (R, t) = \frac{M_\mathrm{disc}}{2\pi R_1 R}T^{-1.5}\mathrm{e}^{-R/(R_1T)},
\end{equation}
where $T=t/t_v+1$, $R_1$ is the characteristic radius of the disc, and $t_v$ is the viscous time-scale at $R_1$, $t_v = R_1^2/(3\nu_1$) \citep{HartmannEtal1998}. After a time $t_0$ gas outside the planet radius starts to dissipate by photoevaporation. Following \citet{AlexanderEtal2006}, we take the photoevaporation rate $\dot{M}_\mathrm{pe}$ proportional to $r_\mathrm{f}^{1/2}$, where $r_\mathrm{f}(t)$ is the radius of the photoevaporation front.
Then we numerically solve for $r_\mathrm{f}(t)$ by mass conservation,
\begin{equation}
    \dot{M}_{\mathrm{pe},i}\left(\frac{r_\mathrm{f}}{r_{\mathrm{f},i}}\right)^{\frac{1}{2}} = 2\pi r_\mathrm{f} \dot{r}_\mathrm{f}\Sigma_v(r_\mathrm{f}, t),
\end{equation}
where $\dot{M}_{\mathrm{pe},i}$ is the mass depletion rate when photoevaporation first carves a gap at $r_{\mathrm{f},i}$.
The default simulation (\Tb{application parameters}) takes the value of $\dot{M}_{\mathrm{pe},i}$ and $r_{\mathrm{f},i}$ from \citet{AlexanderEtal2006} assuming an ionizing flux $\Phi=10^{42}\mathrm{s}^{-1}$ and a disc aspect ratio
\begin{equation}
    h = 0.1\left(\frac{r}{30\ \mathrm{au}}\right)^\frac{1}{4}\left(\frac{\mu}{2.34}\right)^{-\frac{1}{2}} ,
\end{equation}
where $\mu$ is the mean molecular weight.

The gas surface density decreases after the photoevaporation front has passed. We assume the decay of gas surface density \textit{at a fixed point} within the photoevaporation front is exponential. Though the total disk clearing time by photoevaporation is ${\sim}10^5-10^6\ \mathrm{yr}$ \citep{Rosotti2015}, the $e$-folding time-scale $t_\mathrm{pe}$ of gas density at a fixed point is relatively short, $10^3-10^4\ \mathrm{yr}$, estimated from \citet{AlexanderEtal2006i}, \citet{OwenEtal2010} and \citet{GortiEtal2015}. We take $t_\mathrm{pe}=10^4\ \mathrm{yr}$ by default. 
As the planet accretion is insignificant during photoevaporation (\Se{appresult}), changing $t_\mathrm{pe}$ to the lower end of the estimation, $10^3\ \mathrm{yr}$, reduces the amount of accreted pebbles by only $1.4\times 10^{-3} M_\oplus$. 
After photoevaporation, thermal-desorption releases the CO molecules in the dust grains, forming a secondary disc \citep{KrijtEtal2020}. For simplicity, we assume that CO ice constitute a fraction $f_\mathrm{CO}$ of the particles' mass and that it is released to form a secondary CO disc instantaneously once the photoevaporation front has passed the planet.
The mean molecular weight is taken $\mu=2.34$ before photoevaporation, while in the debris disc phase $\mu=28$, consistent with a `shielded secondary disc' dominated by CO \citep{KralEtal2019}. The dispersal time-scale of the $\mathrm{CO}$ gas $t_\mathrm{CO}$ in debris disc is taken to be $2\times 10^6\,\mathrm{yr}$, consistent with the lifetime for which $\mathrm{CO}$ is shielding by carbon as modelled by \citet{KralEtal2019}.

\subsection{Dust model}
\label{sec:app-dust}
The schematic of our simulation set-up is shown in \fg{application-setup}. The dust-to-gas ratio of small dust at the begining of the simulation is $f_\mathrm{dg}$. For simplicity, we fix the dust size $a$ throughout the simulation. 
For the radial drift of the dust, we used a Lagrange smooth particle method similar to \citet{SchoonenbergEtal2018} and simulate the entire system for $t_\mathrm{final}=2\times 10^7\,\mathrm{yr}$.  The Lagrange method integrates the radial motion of `superparticles', each of which represents a group of particles with identical physical properties. When the dust-to-gas ratio in the disc midplane ($Z_\mathrm{mid}$) exceeds $\max \{\mathrm{St}, 1\}$, collective effects reduce the radial motion of dust compared to the gas-rich case (\eq{vdrift}) because the dust accelerates the gas \citep{NakagawaEtal1986}:
\begin{equation}
    \label{eq:v-col}
    v_\mathrm{dr, col} = \frac{2\eta\mathrm{St}}{\mathrm{St}^2+(1+Z_\mathrm{mid})^2}v_\mathrm{k}.
\end{equation}
However, \citet{JiangOrmel2021} argued that considerable amount of dust can `leak' from the edge of a dust ring as there is always a region where $Z_\mathrm{mid}\sim1$.  In our situation, we argue that the planet itself constitutes such an edge, which separates region of low $Z$ (interior to the planet) to region of high $Z$ (exterior to it). This leaking mass flux can be found by maximizing the $Z_\mathrm{mid}$-dependence of the mass flux expression, $\mathcal{M} = 2\pi r \Sigma_d v_\mathrm{dr} = 2\pi C r \Sigma_g Z_\mathrm{mid} v_\mathrm{dr}(Z_\mathrm{mid})h_\mathrm{eff}/h$ where $C$ is a geometrical factor of order unity, $h$ and $h_\mathrm{eff}$ are the gas and pebble aspect ratio, respectively.
This gives $Z_\mathrm{pk}=\sqrt{1+\mathrm{St}^2}$ and 
\begin{equation}
    \mathcal{M}_\mathrm{pk} = 2\pi r C \frac{\sqrt{1+\mathrm{St}^2}-1}{\mathrm{St}} \frac{h_\mathrm{eff}}{h} \eta v_\mathrm{k} \Sigma_g
    \label{eq:Mleak}
\end{equation}
independent of $Z_\mathrm{mid}$. Therefore, the effective velocity of the superparticles corresponding to the peak mass flux reads
\begin{equation}
    v_\mathrm{dr, leak} = \frac{\eta v_\mathrm{k}(\sqrt{1+\mathrm{St}^2}-1)}{Z_\mathrm{mid}};\qquad \left(Z_\mathrm{mid} > Z_\mathrm{pk} \right)
    \label{eq:leak}
\end{equation}
which we use in lieu of \eq{v-col} if $Z_\mathrm{mid}>Z_\mathrm{pk}$.

For $h_\mathrm{eff}$, as particles settle to the disc mid-plane, the aspect ratio of the dust disc is smaller than that of the gas disc, $h_\mathrm{eff}=h\sqrt{\alpha/(\mathrm{St}+\alpha)}$. However, when $h_\mathrm{eff}$ is small enough, Kelvin-Helmholtz instability would operate to stir dust to higher scale height \citep{CuzziEtal1993,YoudinShu2002}. Thus, we take 
\begin{equation}
    h_\mathrm{eff} = \max{\left(h\sqrt{\frac{\alpha}{\mathrm{St}+\alpha}}, h^2 \right)},
    \label{eq:hp}
\end{equation}
where $\alpha$ is the turbulent parameter \citep{ShakuraSunyaev1973}.

\begin{figure}
    \centering
    \includegraphics[width=\columnwidth]{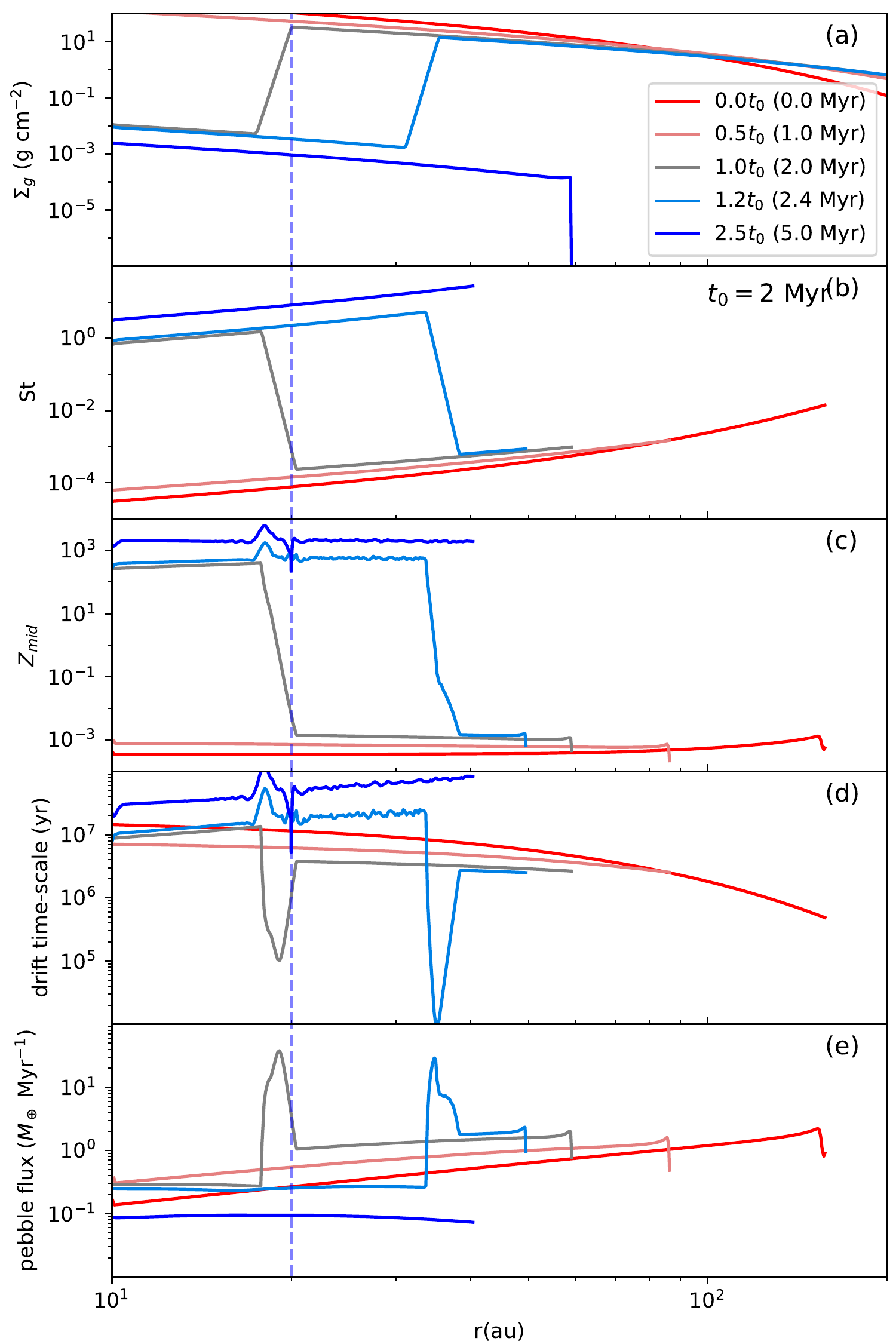}
    \caption{\label{fig:sigma_gas}Evolution of gas and dust properties before, during and after photoevaporation. From top to bottom panels show: gas surface density, Stokes number of the dust particles, midplane dust-to-gas ratio, the drift time-scale and pebble flux for the default model. The time when the photoevaporation front passes the planet's location (vertical dashed line) is $t_0=2\ \mathrm{Myr}$. In the default model, the disc dispersal time-scale is $t_\mathrm{pe}=10^4\ \mathrm{yr}$ and the CO dispersal time-scale is $t_\mathrm{CO}=2\times 10^6\ \mathrm{yr}$.}
\end{figure}

\begin{figure*}
    \centering
    \includegraphics[width=\textwidth]{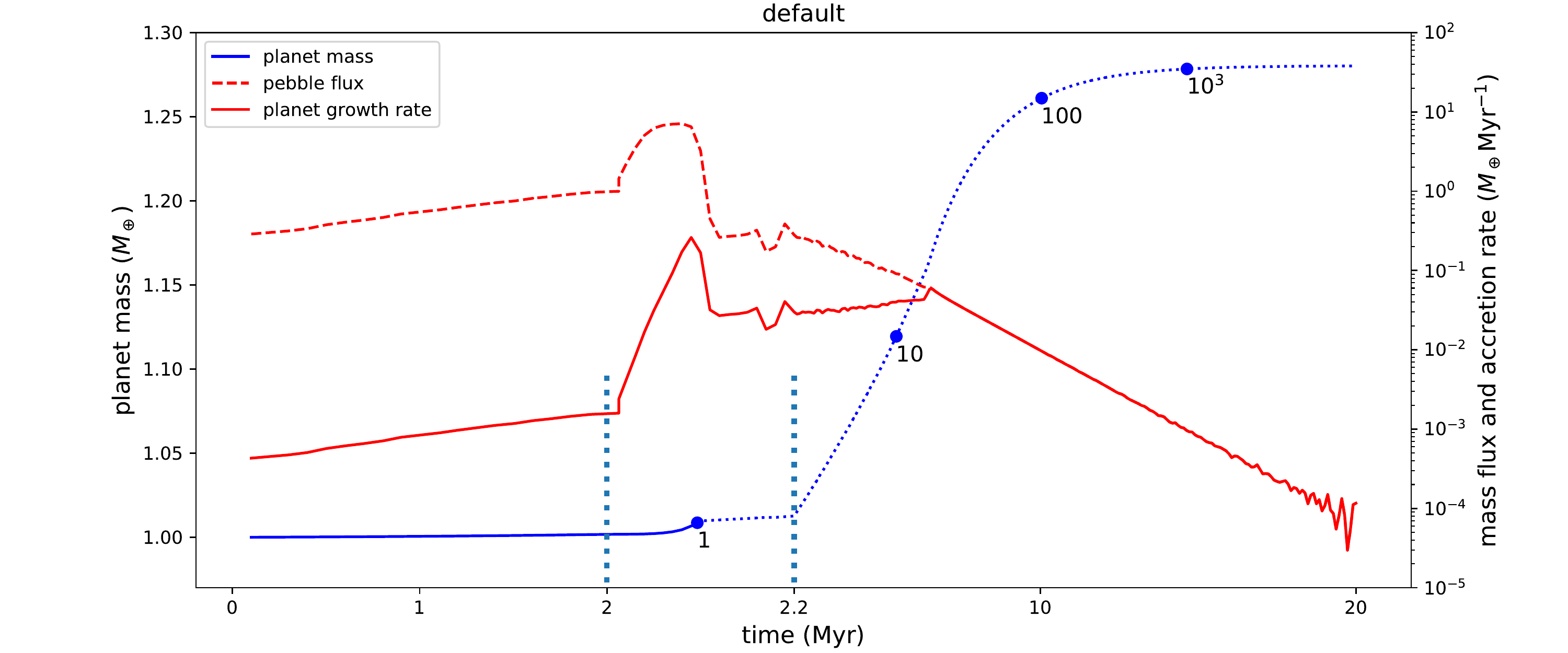}
    \caption{\label{fig:application-default}Planet growth by accreting drifting $30\,\mu\mathrm{m}$ dust particles in a photoevaporating disc. Planet mass (blue line), pebble flux (red dashed line) and planet growth rate (red solid line) are plotted for the default run. The solid part of the blue line indicates acretion by $\mathrm{St}<1$ pebbles while the dotted part indicates $\mathrm{St}>1$. We show when the Stokes number of the dust particles reaches 1, 10, 100, 1000 by blue dots (left to right). 
    The two blue dotted vertical lines indicate $t_0$, the time photoevaporation font passes the planet, and $t_0+20t_\mathrm{pe}$, after which the $\mathrm{H}_2$-rich gas at planet location has been depleted. The x-axis is scaled linearly but the scaling differs in each of the three regions separated by the two vertical dotted lines.}
\end{figure*}

Once the dust superparticle drifts past the planet orbit, its accretion probability is calculated following \tb{ana_sum}. For $\mathrm{St}<1$ particles, we take the 3D limit in \citet{OrmelLiu2018} for the accretion efficiency: 
\begin{equation}
    \epsilon_\mathrm{3D} = 0.39\frac{q_p}{\eta h_\mathrm{eff}}.
\end{equation}
For the accretion efficiency of particles beyond the resonance Stokes number $\mathrm{St}_\mathrm{res}$, we recursively reduce the pebble mass by half until its Stokes number lies in the plateau region, to mimic the fragmentation of pebbles in resonance exposed to high collision velocity \citep{WeidenschillingDavis1985}. The collision frequency of the pebbles is calculated through $t_c = 1/(n\sigma\Delta v)$, taking the mid-plane pebble density at the planet location $3\ \mathrm{Myr}$ after photoevaporation ($t=5\ \mathrm{Myr}$ in \fg{sigma_gas}) and assuming $\Delta v=0.1v_\mathrm{k}$ (\fg{rv}). We found the collisional time-scale to be $~10^{-2}\ \mathrm{yr}$, seven orders of magnitude shorter than the drift time-scale.

\subsection{Default run}
\label{sec:appresult}
\Fg{sigma_gas} shows how the gas profile, dust Stokes number, midplane dust-to-gas ratio, drift time-scale and pebble flux evolves with time for the default model. The drift time-scale is defined as $-r/v_\mathrm{dr}$, where $v_\mathrm{dr}$ is given by \Eqs{vdrift}{leak}. 
The first two lines ($t=0$ and $t=0.5$) show the quantities during the viscous evolution of the protoplanetary disc.  Due to viscous spreading, the inner disc accretes on to the star while the outer disc spreads outward. As a consequence, the gas surface density inside ${\sim}100\,\mathrm{au}$ decreases, together with the radial drift time-scale (the fourth panel) of the $30\ \mu\mathrm{m}$ dust particles.
The photoevaporation does not start until $1.6\ \mathrm{Myr}$ and the photoevaporation front passes the planet location at $t_0=2\ \mathrm{Myr}$ (the third line in each panel). The drift time-scale of the pebbles just inside the photoevaporation front decreases because the Stokes number increases towards unity as the disc surface density decreases.
However the midplane dust-to-gas ratio also increases during disc dispersal, resulting in significant collective effects (see \se{app-dust}) when both $\mathrm{St}$ and $Z_\mathrm{mid}$ exceed unity. For this reason the drift time-scale increases again after reaching a minimum.  
The photoevaporation front evolves from $20\ \mathrm{au}$ to $40\ \mathrm{au}$ within $0.4\ \mathrm{Myr}$ (the third and fourth lines). 
Then at $4\ \mathrm{Myr}$ (not shown in \fg{sigma_gas}), the primordial $\mathrm{H}_2$ disc is cleared from inside-out by photoevaporation, leaving an inner $\mathrm{CO}$ disc of mass $0.71\ M_\oplus$ within $59\ \mathrm{au}$ as the result of outgassing of the icy dust grains.
As the $\mathrm{CO}$ dissipates on a time-scale of 2 Myr (the last two lines), the pebble drift becomes slower with time. After disc dispersal, the mid-plane dust-to-gas ratio and drift time-scale fluctuate. This is attributed to the discretisation effects in our algorithm that computes the dust surface density. However, the pebble mass flux is smooth (last panel), because it reaches an asymptotic limit that is independent of the dust-to-gas ratio $Z_\mathrm{mid}$ [see \eq{leak}, for more details].

The mass accreted by the planet in our default simulation is shown in \fg{application-default} and in the first row of \Tb{application result}. The disc surface density decreases during the first stage. Viscous spreading increasing the pebbles' Stokes number, resulting in faster drift and therefore a higher pebble flux. When photoevaporation starts, the disc dissipates from inside-out, during which the Stokes number of pebbles changes from $\mathrm{St}\ll 1$ to $\mathrm{St}\gg 1$. 
The pebbles' drift velocity peaks $0.1\ \mathrm{Myr}$ after the photoevaporation front passes. The pebble accretion efficiency also increases, because the pressure gradient parameter decreases with increasing mean molecular weight after dispersal. Therefore, the planet growth rate peaks at ${\sim}0.1\,M_\oplus\,\mathrm{yr}^{-1}$ after $2.1\ \mathrm{Myr}$.
Thereafter, the mid-plane dust-to-gas ratio $Z_\mathrm{mid}$ exceeds unity and the particle drift is suppressed due to collective effect. As a result of the short dispersal time-scale ($10^4\ \mathrm{yr}$), the planet does not grow much during the dispersal phase. 

After the quick disc dispersal, the $\mathrm{St}\gg1$ pebbles remaining outside of the planet's orbit drift slowly. The pebble flux decays on the time-scale over which the secondary CO disc disperses ($t_\mathrm{CO}$). However, these pebbles are accreted in the ballistic rise regime, for which $\epsilon(\mathrm{St})$ increases. Despite a decreasing mass flux, the pebble accretion rate still increases until $4.5\ \mathrm{Myr}$ after dispersal. 
As the value of the Stokes number at $6.5\ \mathrm{Myr}$ is $\mathrm{St} = 17$, planet growth is characterised by a high accretion efficiency (the plateau region) after which, the planet accretes pebbles at 100 per cent probability. Because the pebble mass flux decreases in the debris disc phase, the growth of the planet slows down accordingly. By $t=10\,\mathrm{yr}$ its mass asymptotes out at $\approx{1.25}\,M_\oplus$, 20 per cent of which are accreted after disc dispersal. 

\begin{table}
    \small
    \caption{Planet mass accreted in each Stokes number regime. The second column shows the parameters used in each run. Except for run \texttt{default} where we take parameters in \Tb{application parameters}, we change one parameter in each run and leave others unchanged. Column $m_{\mathrm{St}<1}$, $m_\mathrm{rise}$ and $m_\mathrm{pla}$ denote dust mass accreted as $\mathrm{St}<1$ pebbles, pebbles in ballistic rise stage and in plateau stage.}
    \label{tab:application result}
    \centering
    \begin{tabular}{c c  l l l}
    \hline
    name & parameters & $m_{\mathrm{St}<1}$ & $m_\mathrm{rise}$ & $m_\mathrm{pla}$\\
    & & $M_\oplus$ & $M_\oplus$ & $M_\oplus$ \\
    \hline
    \texttt{default} &  \Tb{application parameters} & $0.0097$ & $0.15$ & $0.12$ \\
    \texttt{lowZ0} & $f_\mathrm{dg} = 0.0001$ & $0.0029$ & $0.079$ & $0.10$\\
    \texttt{highmp} & $M_p = 2$ & $0.019$ & $0.24$ & $0.17$\\
    \texttt{lowfCO} & $f_\mathrm{CO} = 0.05$ & $0.0090$ & $0.11$ & $0.12$\\
    \texttt{small} & $a = 10\ \mu \mathrm{m}$ & $0.062$ & $0.066$ & $0.040$\\
    \texttt{shorttCO} & $t_\mathrm{CO} = 1\mathrm{Myr}$ & $0.0096$ & $0.081$ & $0.060$\\
    \hline
    \end{tabular}
\end{table}

\begin{figure}
    \centering
    \includegraphics[width=\columnwidth]{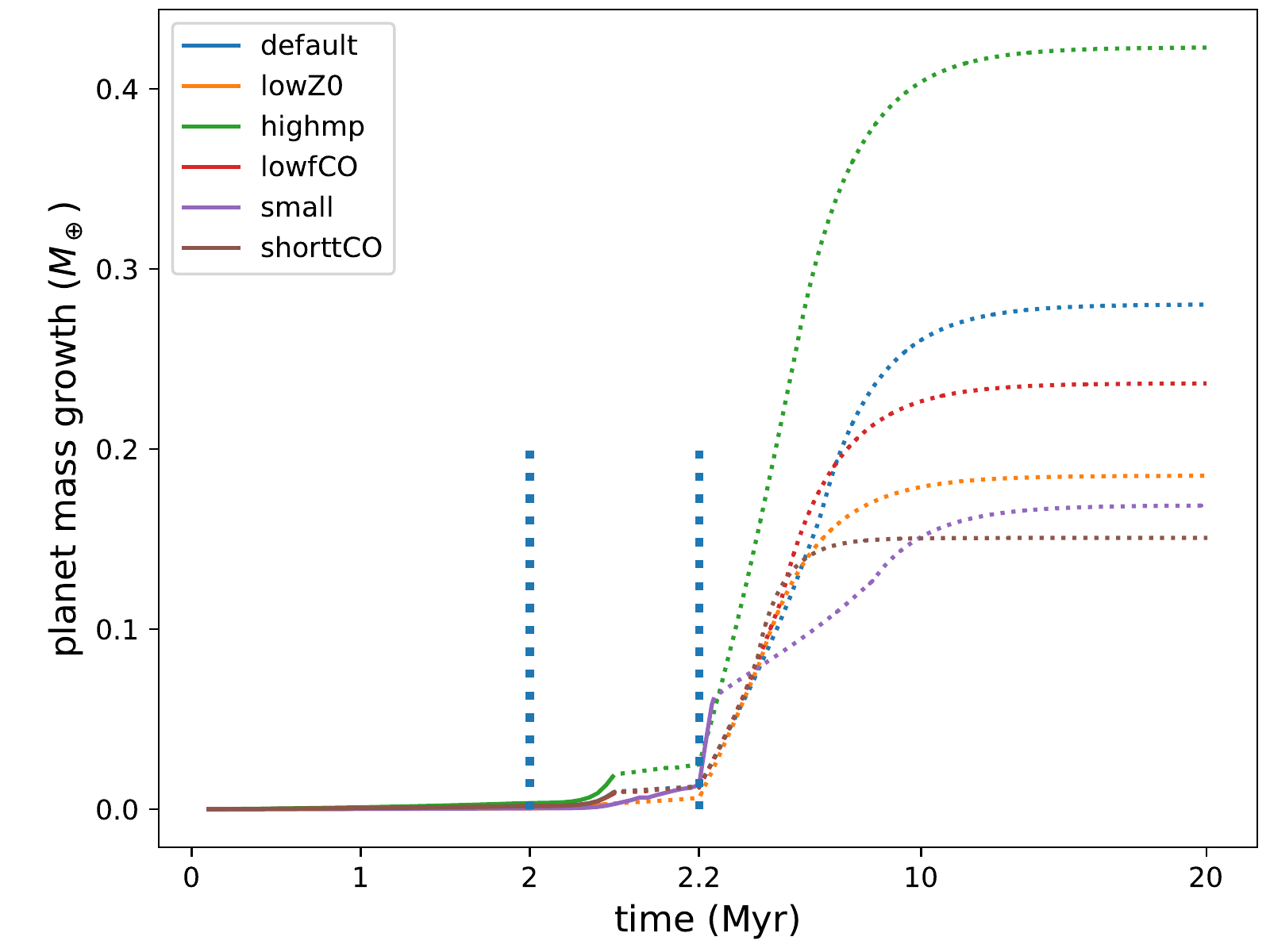}
    \caption{\label{fig:application-all}Planet mass growth vs.\ time for different models. See \Tb{application result} for the parameters changed in each run. Solid and dotted parts, and vertical dotted lines have the same meaning as \Fg{application-default}. The scaling of the x axis is the same as \Fg{application-default}.}
\end{figure}

\subsection{Parameter variation}
We explore the influence of the following model parameters: initial dust-to-gas ratio $f_\mathrm{dg}$, planet initial mass $M_{p0}$, CO mass fraction of dust $f_\mathrm{CO}$, dust size $a$ and CO gas dispersal rate $t_\mathrm{CO}$. For each run, one of these parameters is varied compared to the default model, while other parameters are left unchanged. 
\Tb{application result} lists the mass accreted in different Stokes number regime, and \Fg{application-all} shows the results of parameter studies graphically. 
The general features -- negligible growth by small Stokes number pebbles and sustained growth in the debris disc phase --  are similar among different models. 

The mass accretion rate (the slope in \Fg{application-all}) in the ballistic rise regime is insensitive to $Z_0$, $f_\mathrm{CO}$ and $t_\mathrm{CO}$. This is because the dependence of the mass flux [\Eq{Mleak}] and the accretion probability [\Eqs{rise}{St_Ep}] on the gas surface density cancel: $\dot{M}_p = \mathcal{M}_\mathrm{pk}\epsilon_\mathrm{bal,1} \mathrm{St} \propto \Sigma_g \mathrm{St} \propto (\Sigma_g)^0$ as $\mathrm{St}\propto 1/\Sigma_g$ in the Epstein drag limit. As the mass accretion rate neither relies on gas nor dust surface density, 
for run \texttt{lowZ0}, \texttt{lowfCO} and \texttt{shorttCO}, the mass accreted in the ballistic rise regime only depends on the time when accretion plateau regime starts. In \texttt{lowZ0}, we decrease $f_\mathrm{dg}$ by a factor of 3, meaning that an equivalently lower amount of CO is released after disc dispersal. Consequently the transition between the ballistic rise regime to the plateau regime already happens at $4.3\ \mathrm{Myr}$, $2.1\ \mathrm{Myr}$ earlier than in the default run, 
which results in $0.07M_\oplus$ less dust accreted in the ballistic rise regime (see \Tb{application result}). The same explanation can be invoked to run \texttt{lowfCO}, where we decrease the CO mass fraction in the dust particles by half. In this test, the mass of CO in the debris disc phase is decreased by half, and the planet accretes $0.11\ M_\oplus$ in the ballistic rise regime.
When we decrease the CO life time $t_\mathrm{CO}$ to $10^6\ \mathrm{yr}$ (run \texttt{shorttCO}), the Stokes number increased 2 times faster, for which reason the planet accreted roughly half of dust -- $0.081M_\oplus$ than the default run. 
On the other hand, higher planet mass allow the planet to catch the pebble with higher probability. In run \texttt{highmp}, the planet accreted $0.09M_\oplus$ more in the ballistic rise regime.

We combine the discussion for the plateau regime and resonance regime, because to mimic collisional cascade, we continuously cut the pebble mass by half until the accretion efficiency is in the plateau regime. In these regimes, the accretion probability is nearly 100 per cent. Therefore the accretion rate only depend on the pebble drift flux.
As shown in \eq{Mleak}, the leaking flux is only a function of CO surface density at the beginning of plateau region and the CO depletion time-scale $t_\mathrm{CO}$. Note that the CO surface density is one-to-one related to the Stokes number [\eq{St_Ep}]. Therefore, the mass flux in the plateau regime is only a function of $\mathrm{St}_\mathrm{plat}$ and $t_\mathrm{CO}$. For \texttt{lowZ0} and \texttt{lowfCO}, $\mathrm{St}_\mathrm{plat}$ [\eq{St_pla}] is identical to the default case, so the mass accreted in the plateau regime and rising regime in these runs is similar to that of the default run.
In run \texttt{shorttCO}, the mass accreted in these stages is about half of the default run, because the CO disc disperses two times faster. In run \texttt{highmp}, $\mathrm{St}_\mathrm{pla} = 11$, less than the default value $17$. Hence the gas surface density at the onset of the plateau regime is higher, resulting in larger amount of pebbles accreted.

Combining the ballistic rise stage and plateau stage, we see that the dust mass accreted is not linearly related to the initial dust-to-gas ratio and CO mass fraction in the dust. If we decrease $Z_0$ by a factor of 3, planet growth only decreases by 34 per cent. This suggests that even a small reservoir of leftover dust can already make a significant contribution to the growth of the planet after disc dispersal.

One exception to the above discussion is run \texttt{small}, where we decrease the particle size by a factor of 3 to $10\ \mu\mathrm{m}$. The Stokes number only exceeds unity after $2.7\ \mathrm{Myr}$ due to the smaller size. Therefore, the dust is accreted more efficiently when they are aerodynamically small ($\mathrm{St}<1$). 
On the other hand, for the same St, the gas surface density is lower, resulting in a lower leak pebble flux [\eq{Mleak}] and an accretion rate that is a factor of three less in the ballistic rise regime. In this run, the planet only accretes $0.066M_\oplus$ in the rising stage and $0.04M_\oplus$ in the plateau and resonance stages.
Note that increasing the dust grain size would not result in a larger planet mass, as the dust particle would already have drifted interior to the planet location before the disc dispersal.

\subsection{Caveats}
The above numerical experiments imply that accretion of large Stokes number pebbles can be important for the late growth of planets, and perhaps even the dominant factor, especially when the CO lifetime is long in the disc. None the less, several caveats present in our models and assumptions here must be addressed to solidify this scenario. First, the disc model is greatly parameterised. For example, we assume the gas surface density experiences exponential decay after the photoevaporation front has passed. However, the disc dispersal rate $\dot{\Sigma}(r, t)$ relies on the details of photoevaporation model, which is influenced by uncertain factors like soft X-ray intensity \citep{ErcolanoEtal2021, SellekEtal2022}, stellar mass \citep{PicognaEtal2021} and disc opacity \citep{NakataniEtal2021}. Coupling our accretion model with accurate disc model including XEUV physics and radiation transfer \citep{PicognaEtal2019} will provide a better description for the photoevaporation rate. Besides photoevaporation, disc winds can also drive mass loss in the disc \citep{SuzukiInutsuka2009, ArmitageEtal2013, BaiStone2013}, which is left out of consideration in this work. 

Second, the assumption that the Bondi radius equals the planet physical radius in the debris disc phase should be treated with caution. As the gas in the disc is photoevaporated, it is not likely that the planet can maintain hydrostatic balance with the dilute CO gas. More realistically, the reduction in the pressure scaleheight due to the high-$\mu$ gas will enable the planet to simply consume all the CO gas in its feeding zone. How the dust particles respond to this environment is an open question. 

Notwithstanding these caveats, as long as there is considerable amounts of dust and gas left in debris discs, as is hinted observationally (\se{application-gas}), the possibility of planet growth by accreting large Stokes number pebbles is conceivable and an scenario worth further investigation.
\begin{figure}
    \includegraphics[width=\columnwidth]{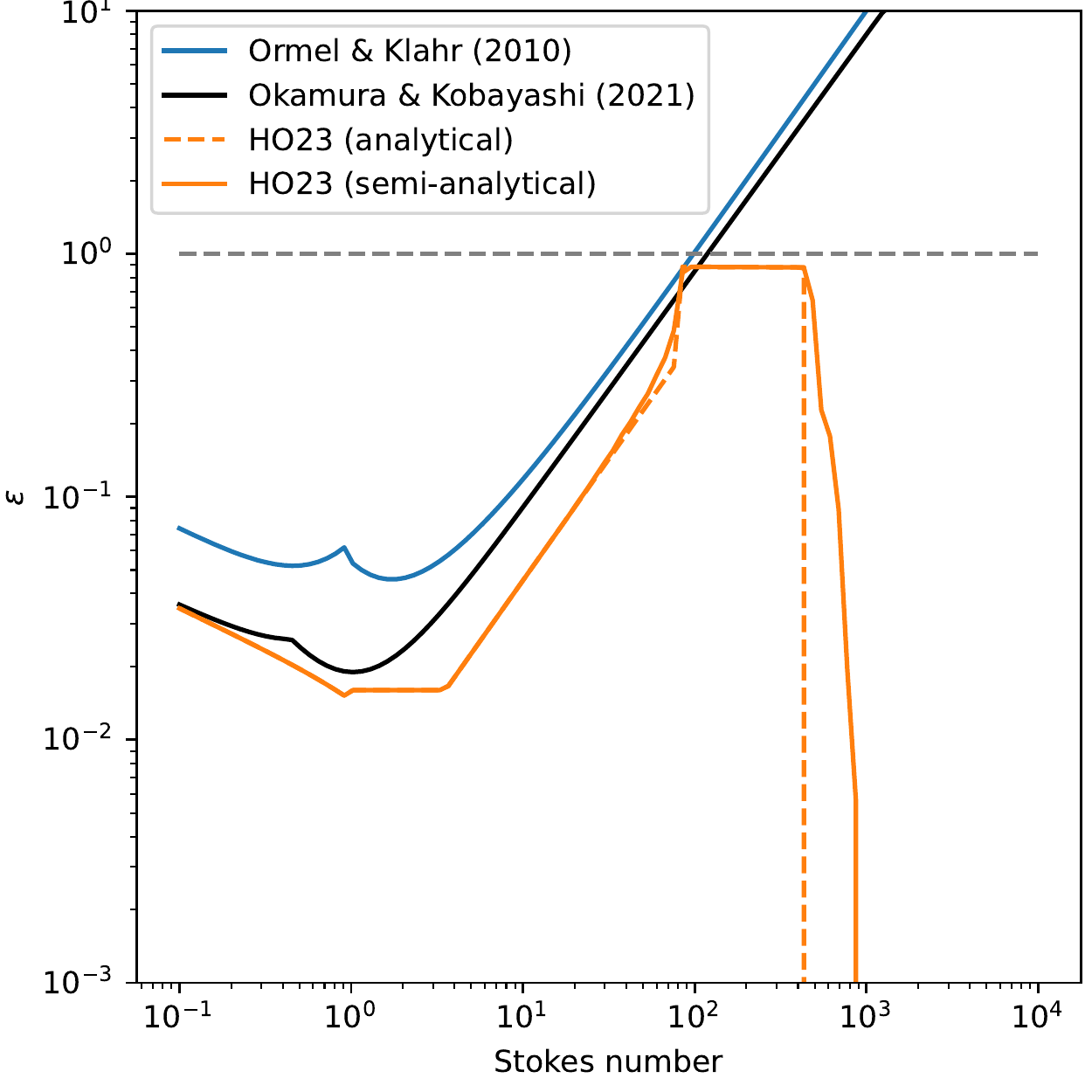}
    \caption{\label{fig:OKvsHO} Pebble accretion efficiency of \citet{OrmelKlahr2010}, \citet{OkamuraKobayashi2021} and our work. The parameters are the same as the default in \Tb{parameters}, i.e. $q=3\times 10^{-6}$, $\eta = 3\times 10^{-3}$, $R_p=R_\mathrm{b}$.
    }
\end{figure}

\section{Discussion}
\label{sec:discussion}
\subsection{Comparison with previous works}
This work continues a series of papers about the PA efficiency. 
\citet{OrmelKlahr2010} (OK10 hereafter) conducted 2D numerical integrations of a pebble's motion in a local frame and obtained expressions for its accretion rate in a laminar disc. They expressed the accretion rate in terms of Hill units, $\dot{M}=r_\mathrm{h}^2 \Sigma \Omega P_\mathrm{col}$, which is related to the pebble accretion efficiency ($\epsilon$) in the following way:
\begin{equation}
    \label{eq:pcol-to-eps}
    P_\mathrm{col} = \frac{2\pi r v_r \epsilon}{r_\mathrm{h}^2 \Omega}
    = \frac{4\pi \eta}{\mathrm{St} {r'}_\mathrm{h}^2} \epsilon \qquad (\mathrm{St}\gg1)
\end{equation}
Focusing on the $\mathrm{St<1}$ limit, \citet{LiuOrmel2018} and \citet{OrmelLiu2018} gave expressions for the efficiency of settling and ballistic hits in the 2D and 3D limits, respectively. Like in this study they use a global setup to directly obtain $\epsilon$. 
Finally, \citet{OkamuraKobayashi2021} (OK21 hereafter) included planet-induced gas flow into the accretion of solid objects. They use hydrodynamic simulations near the planet region to yield the density and velocity profile of the gas. They consider a local geometry and conduct integrations for Stokes numbers in the range of $10^{-2}$ to $10^{13}$.

\Fg{OKvsHO} summarizes the expressions from these works against our (semi-)analytical expressions, where we focus on the large Stokes particle limit.
We use the default parameter in \Tb{parameters} and take the planet radius to be the Bondi radius.
In the regime before the plateau region ($\mathrm{St<100}$), it shows that the OK10 expressions and, to a lesser extent, the OK21 expressions tend to overestimate $\epsilon$. 
This is mainly because these local simulations consider pebbles coming form both interior and exterior to the planet orbit. However, in the global frame pebbles with Stokes number in the ballistic rise stage ($1\lesssim\mathrm{St}\lesssim100$ in this case) are actually drifting so fast that they seldomly enter the Hill sphere from the inner side. For this reason, our $P_\mathrm{global}$ is reduced by a factor of 2 compared to these two previous works.

In the plateau region ($100\lesssim \mathrm{St}\lesssim 500$) $\epsilon$ for our global simulations hovers around unity, while in the local simulations of OK10 and OK21 the accretion efficiency would exceed unity. The meaning of $\epsilon>1$ here is that the accretion time-scale is faster than the pebble supply time-scale, so that the radial drift would be unable to replenish the pebbles. Physically, the pebble surface density around the planet's orbit would decrease, and the pebble accretion rate is limited by the pebble drift flux, which is consistent with our result.

The largest difference appears at the resonance stage ($\mathrm{St}\gtrsim 500$). In our simulation, the pebbles are prevented from accretion by the planet `kick'. So the shearing box initial condition in OK21 is not reachable once the planet has accreted all the large pebbles near its orbit. By the above comparison, we reach the conclusion that to consider the accretion of $\mathrm{St}>1$ pebbles, one better traces the pebble's trajectory in the global frame. 
Though in our simulation a single pebble will be trapped in resonance when the Stokes number is larger, the overlap of resonance orbits when $j>2$ and probabilistic trapping (see \se{result}) make the collision and fragmentation possible. This is in agreement with \citet{WeidenschillingDavis1985}, who estimated that a swarm of planetesimals trapped in $j>2$ resonance near an Earth-like planet collides with velocity ${\sim}1\ \mathrm{km}\ \mathrm{s}^{-1}$. Therefore they argued that the planetsimals may fragment to smaller sizes to be accreted by the planet, justifying the accretion probability of resonance pebbles used in \se{application}.

\subsection{Limitations}
We state the limitations of our model here. First, we assumed a laminar discs with constant pressure gradient parameter, ignoring any disc substructure, caused by interaction with the planet \citep{LinPapaloizou1993}. Specifically, once the planet is massive enough, i.e. $q\sim h^3$, it opens a gap near its orbit by tidal interaction with the disc. Small pebbles ($\mathrm{St}<1$) can be trapped in the pressure maximum at the outer edge of the pressure bump, even though they are located close to the planet, at a distance of ${\sim}r_\mathrm{h}$ \citep[e.g.]{ZhuEtal2014}. However, aerodynamically large pebbles will avoid trapping, because the time to respond to the pressure bump (the pebble stopping time) is longer than the gravitational interaction with the planet (${\sim}\Omega^{-1}$).  Hence, presssure bumps cannot prevent the accretion of aerodynamically large pebbles. 
Arguably more critical to the accretion of large pebbles is the nature of the atmosphere of these large planets. They will presumably have collapsed, perhaps into a circumplanetary discs, in which case the accretion efficiency will be different from our study.

Second, the planet gravitational perturbation induces the spiral density wave in the gaseous disc, which may influence the orbital decay of the large pebbles \citep{MutoInutsuka2009}. As the large pebbles are less influenced by the gas motion, we do not expect spiral density wave to dominate their motion. Calculation by \citet{MutoInutsuka2009} show that the orbital decay rate due to the gas radial pressure gradient exceeds that of the spiral density wave by two orders of magnitude, which justifies our assumption of ignoring planet-disc interaction.

Third, in each of our simulation, we fixed the parameters as in \Tb{parameters}. However, the disc is evolving viscously \citep{Lynden-BellPringle1974} or through disc winds \citep{ArmitageEtal2013, BaiStone2013, TaboneEtal2022}. If the disc surface density changes over the pebble's interaction time-scale with the planet, our assumption of fixed Stokes number is no longer justified. However, in the plateau stage, most of the simulations complete within $2000$ planet orbits. Therefore, our constant parameter assumption is acceptable, except for the outer-most disc regions.

Finally, we integrate the pebble's orbit in the 2D plane. This is valid when the vertical scale height of the pebbles (\eq{hp}) is smaller than the Hill radius of the planet. If we take the turbulent parameter $\alpha=0.01$, the orbit of large pebbles ($\mathrm{St}>1$) in parameter space studied in this work (\Tb{parameters}) can be approximated by 2D motion. However, when the disk scale height is higher (e.g. when $h>0.1$) or the planet mass is smaller (e.g. $q<2\times 10^{-7}$), 3D effects may become important, as pebbles may leak through the planet's position.
\section{Conclusions} \label{sec:conclusion}

In this paper we performed 2D orbital integration of $\mathrm{St}>1$ pebbles in a global frame, investigating the accretion potential for these particles. For a planet on a fixed Keplerian orbit, we calculated the probability for these pebbles to either settle towards the planet, like with $\mathrm{St}<1$ pebbles, or to hit the protoplanet's (atmospheric) radius ballisticlly. A semi-analytical model (\se{semiana}) and a fully analytical fit (\Tb{ana_sum}) are provided to calculate the accretion efficiency -- that is, the probability of the pebble to be accreted by the planet rather than it to drift inwards or be trapped in resonance. We found that our expressions fit the numerical results within a factor of 1.5 in most cases. Then, we assessed the importance of pebble accretion in the large Stokes number limit to a planet in a dispersing protoplanetary disc and then a debris disc, to study the characteristics of such late stage accretion. 

The main conclusions are as follows:
   \begin{enumerate}
       \item Unlike aerodynamically small pebbles, large Stokes number pebbles are primarily accreted by the planet ballistically, relying on a surface, like in planetesimal accretion.  The existence of a dense and extended atmosphere will render the effective capture radius much greater than the physical radius, which, together with the reduced drift motions greatly boosts the pebble accretion efficiency. 
       \item Large Stokes number accretion is characterised by the the accretion plateau -- referring to the range in Stokes numbers where the accretion probability can almost reach 100 per cent ($70<\mathrm{St}<400$ in \fg{default}). Pebbles whose Stokes number exceeds this range are trapped in resonance while lower Stokes number particles become dominated by drift motions. With increasing planet-to-star mass ratio $q$ the plateau region shifts to lower Stokes numbers. On the other hand, increasing the pressure gradient parameter $\eta$ shifts the accretion plateau to larger Stokes number. 
       \item When the Stokes number is slightly above the plateau-resonance boundary, the pebbles will not be trapped in the resonance orbits with definite j, because different j:j+1 mean motion resonance orbits tend to overlap with each other when j is high. 
           This effect will bring large relative velocity when pebbles collide, likely leading to their fragmentation. After fragmenting into smaller pieces, pebbles end up with a smaller Stokes number, so they can be accreted with high efficiency in the plateau region. 
       \item The accretion behaviour of a pebble is well described by its radial drift during a synodical time and the planet gravitational perturbation during encounter.  Our semi-analytical method based on such repeated drift-and-kicks predicts the resonant Stokes number well (see \se{semiana}). We also developed analytical fits to the simulation result (\se{analytical}), which show a factor of 1.5 agreement with simulation in most regions of the parameter space.
       \item When the primordial gas in the disc disperses due to photoevaporation, the Stokes number of of small (${\sim}10\,\mu\mathrm{m}$) dust particles will exceed unity as the H/He gas is replaced by a more tenuous CO-rich disc. The accretion plateau renders these particles an effective source for planet growth. We find that an Earth-mass planet can accrete $20$ per cent of its total mass from these large Stokes dusts particles in the debris disc. The amount of large pebble accreted through this process depends on the particle size and the lifetime of the debris disc gas, but is found to be less sensitive to the mass fraction of small population dust and the CO ice mass fraction of the dust.  Accretion of aerodynamically big pebbles provides a continuous pathway to grow planets in the debris disc.
   \end{enumerate}

\section*{Acknowledgements}
The authors appreciate the thoughtful comments of the referee, Keiji Ohtsuki. We also acknowledge comments by Xuening Bai, Yixian Chen, Hiroshi Kobayashi, Michiel Lambrechts, Beibei Liu. HH acknowledges support from a LinBridge fellowship. 
CWO acknowledges support by the National Natural Science Foundation of China (grant no. 12250610189).

\section*{Data Availability}
The data underlying this article will be shared on reasonable requests to the corresponding author.
The script of our analytical prediction for accretion efficiency is available at \url{https://github.com/chrisormel/astroscripts/tree/main/papers}.



\bibliographystyle{mnras}
\bibliography{ads} 




\appendix
\section{Results for Stokes Drag Law}\label{sec:appendix}

\begin{figure*}
    \includegraphics[width=\textwidth]{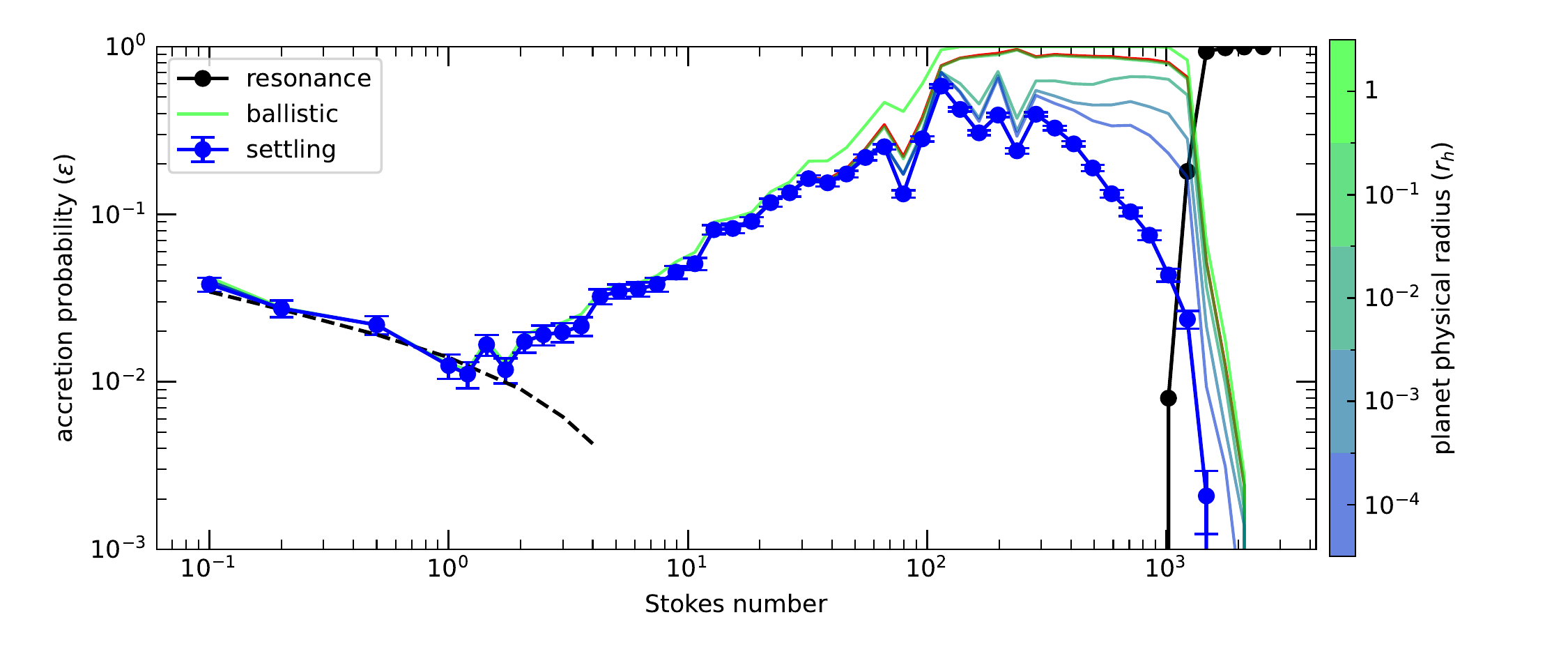}
    \caption{Pebble accretion rate for default parameters, but with Stokes gas drag law. The x axis is the Stokes number consistent with drift-induced relative velocity. The meanings of each line is the same as \Fg{default}.}
    \label{fig:app-default}
\end{figure*}

Here we consider the gas drag in Stokes regime onto particles with sizes $s>9\lambda/4$:
\begin{equation}
    F_{\mathrm{drag, St}} = \frac{1}{2} C_D \pi s^2 \rho v^2.
    \label{eq:St_St}
\end{equation}
The coefficient $C_D$ depends on the Reynolds number $Re=2sv/\nu$, where $\nu$ is the molecular viscosity,
\begin{equation}
    C_D = 
    \begin{cases}
        \displaystyle
        \frac{24}{Re} & Re<54.5; \quad \textrm{(linear)} \\
        0.44 & Re\geq 54.5. \quad \textrm{(quadratic)}
    \end{cases}
    \label{eq:drag_coe}
\end{equation}
Here we take two regimes, for simplicity, ignoring an intermediate expression \citep{Rafikov2004}. We refer to these regimes as `linear Stokes' and `quadratic Stokes', respectively. 
Thus, in the linear Stokes regime $\mathrm{St}$ is independent of the relative velocity between pebble and gas, whereas in the quadratic Stokes regime, $\mathrm{St}$ becomes inversely propotional to relative velocity $v$. 
From the definition of the Reynolds number and \eq{drag_coe}, the linear Stokes regime and the quadratic Stokes regime are separated by a relative velocity threshold of $v_\mathrm{crit} = 27.2\nu/s$.
Adopting the gas surface density at 1\,au in the Minimum Mass Solar Nebular \citep{Hayashi1981} and assuming a disc aspect ratio $h=0.05$ and pebble internal density $\rho_\bullet = 1\ \mathrm{g}\ \mathrm{cm}^{-3}$, we obtain $v_\mathrm{crit} = 4\times 10^{-3}v_\mathrm{k}/\sqrt{\mathrm{St}_\mathrm{lin}/10}$, where $\mathrm{St}_\mathrm{lin}$ is the Stokes number in linear Stokes drag regime.

\Fg{app-default} shows the accretion rate of pebbles undergoing Stokes drag law, with default parameters (cf. \Fg{default}). Stokes drag is more likely at higher gas density, i.e., in the inner disc regions. In the Stokes drag regime, the Stokes number is not constant in time due to its dependence on the relative velocity between the pebble and the gas [\eq{St_St}]. 
In \Fg{app-default} the Stokes number listed on the $x$-axis corresponds to the initial Stokes number of the particles, unperturbed by the planet.
Capture by settling dominates the accretion rate for $\mathrm{St}<200$, because when the pebble is inside the Hill sphere, high relative velocities with the gas will increase the drag force quadratically, instead of linearly for Epstein drag. 
Gas drag therefore dissipates the pebble's kinetic energy more efficiently, rendering the pebble more likely to be captured by settling. 
Apart from this, the general trends of pebble accretion in the Stokes drag regime remain similar to the Epstein regime: here, too, we discern a rising ($1<\mathrm{St}<100$), plateau ($100<\mathrm{St}<1000$) and resonance regime ($\mathrm{St}>1000$). 
This is not surprising as the motion of pebbles outside the Hill radius are well approximated by the `drift and kick' model (see \se{semiana}), during which relative velocities remain modest and Stokes numbers changes due to the Stokes drag law is negligible.

\begin{figure}
    \includegraphics[width=0.5\textwidth]{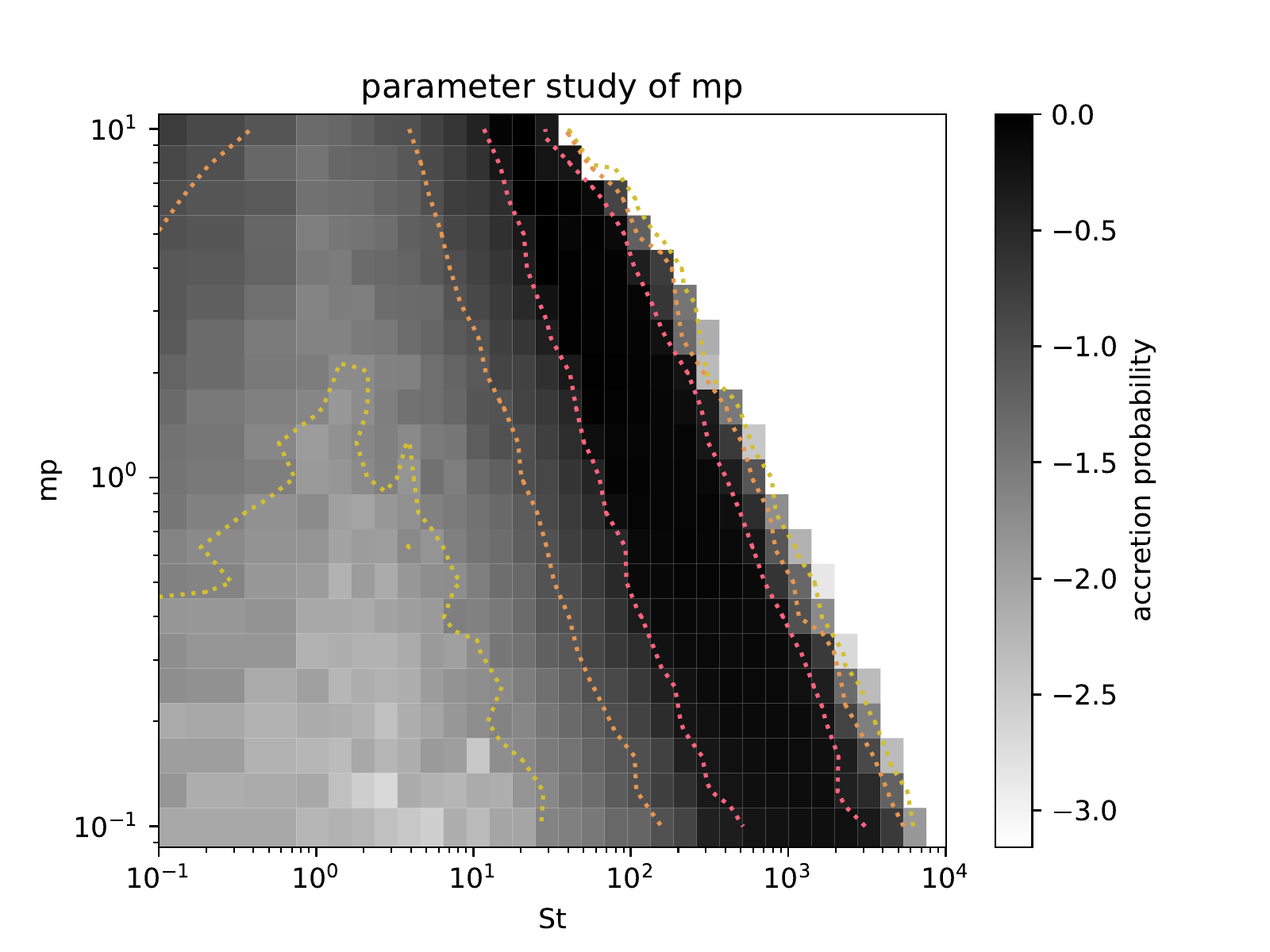}
    \caption{Same as the left panel of \Fg{q}, but for the Stokes drag law.}
    \label{fig:app-q}
\end{figure}

\begin{figure}
    \includegraphics[width=0.5\textwidth]{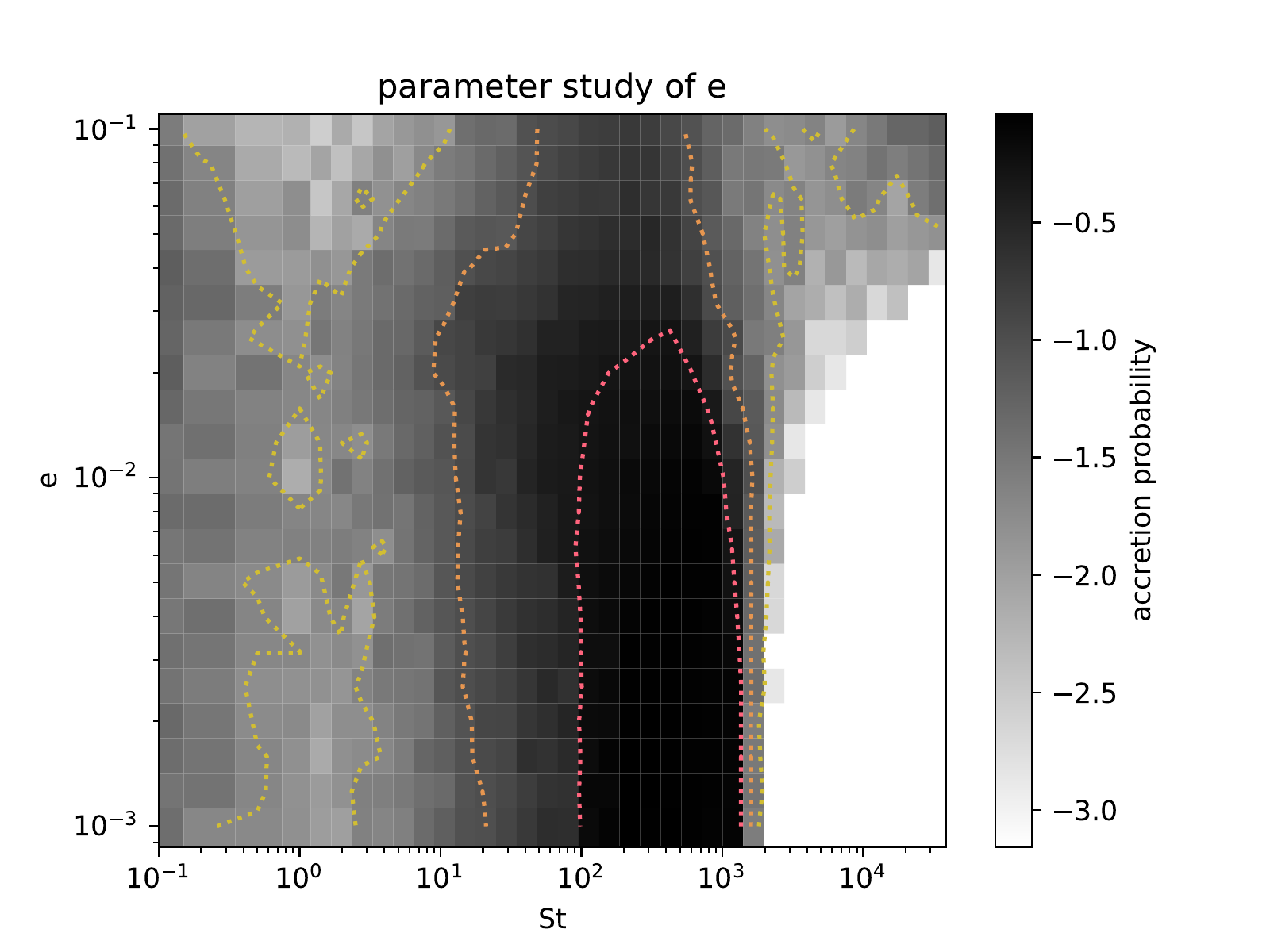}
    \caption{Same as \Fg{e}, but for the Stokes drag law.}
    \label{fig:app-e}
\end{figure}

For reference, \Fg{app-eta}, \ref{fig:app-q} and \ref{fig:app-e} show the results for the parameter study of $\eta$, $q$ and $e_p$. The accretion probability bears little difference with the Epstein drag case, especially for the parameters study of $\eta$ and $q$.
When the planet eccentricity exceeds $0.03$, we found that the pebbles under Stokes drag law are more likely to be trapped into the resonance orbit of a eccentric planet. Thus, accretion rate becomes lower for $\mathrm{St}>100$ compared to Epstein regime. 

\begin{figure}
    \includegraphics[width=0.5\textwidth]{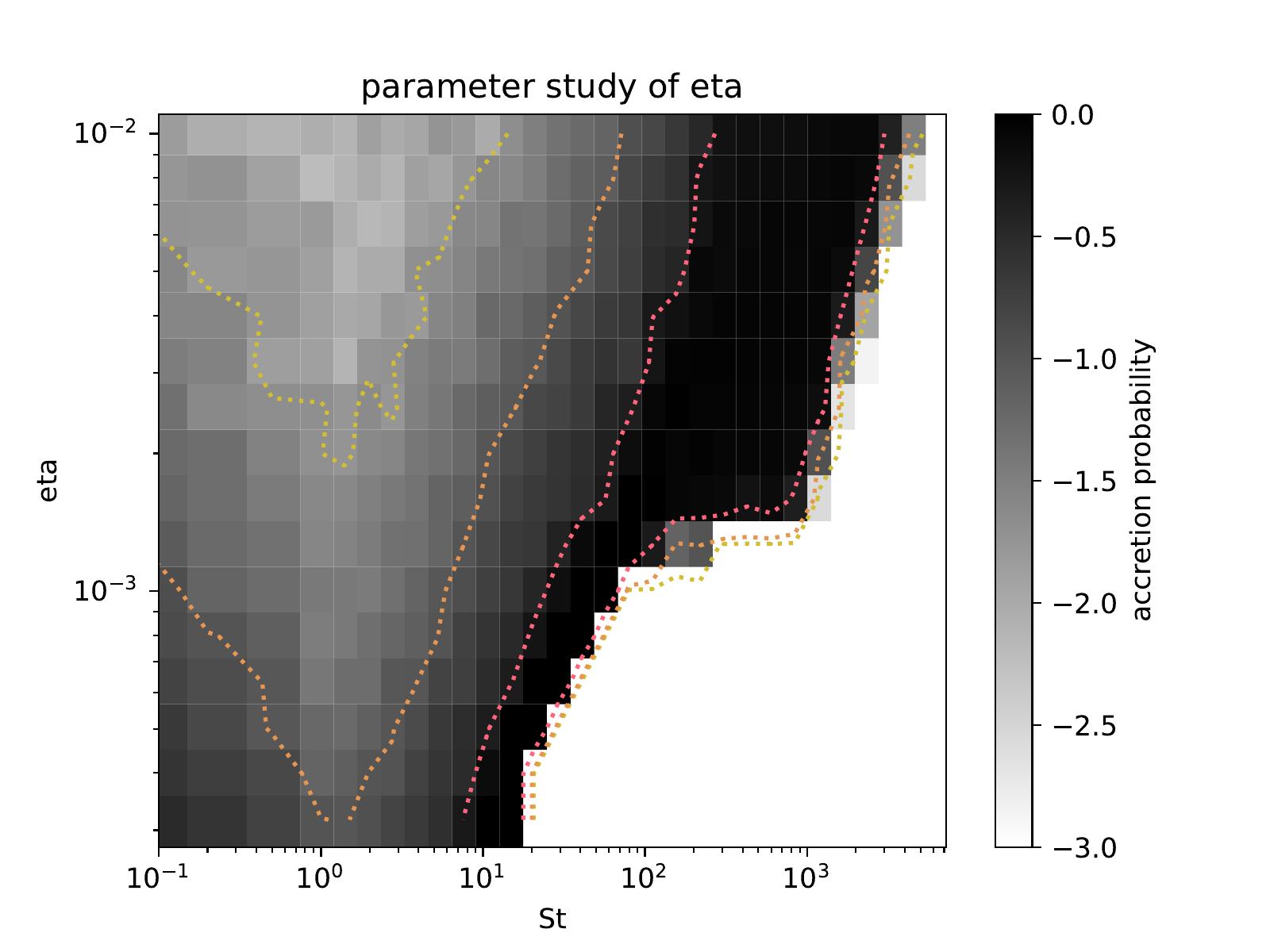}
    \caption{Same as the left panel of \Fg{eta}, but for the Stokes drag law.}
    \label{fig:app-eta}
\end{figure}

\bsp	
\label{lastpage}
\end{document}